\documentclass[a4paper,11pt,fleqn]{article}
\usepackage{amsfonts,times,epsfig}

\newlength{\dinwidth}
\newlength{\dinmargin}
\def\nn{\nonumber}
\def\non{\nonumber\\}
\def\be{\begin{equation}}
\def\ee{\end{equation}}
\def\ben{\begin{displaymath}}
\def\een{\end{displaymath}}
\def\ba{\begin{eqnarray}}
\def\ea{\end{eqnarray}}

\def\a{\alpha}
\def\b{\beta}

\def\d{\delta}
\def\e{\varepsilon}

\def\g{\gamma}
\def\G{\Gamma}

\def\l{\lambda}
\def\L{\Lambda}
\def\m{\mu}
\def\n{\nu}
\def\O{\Omega}
\def\o{\omega}

\def\r{\rho}
\def\s{\sigma}

\def\t{\tau}
\def\th{\theta}

\def\x{\xi}

\def\e{\epsilon}
\def\ps{{\pi_\sigma}}
\def\pr{{\pi_\rho}}

\def\cC{{\cal C}}
\def\cM{{\cal M}}
\def\E{{\cal E}}

\def\cL{{\cal L}}

\def\cV{{\cal V}}
\def\cP{{\cal P}}

\def\rt{{\tilde{\r}}}

\def\R{\mathbb{R}}

\def\E{E_{8(+8)}}
\def\eee{{\mathfrak{e}_8}}
\def\eeee{{\mathfrak{e}_9}}
\def\EE{E_{8(+8)}}
\def\9{E_{9(+9)}}
\def\0{E_{10}}
\def\SO{{SO}(16)}
\def\so{{\mathfrak{so}(16)}}

\def\la{\label}
\def\ci{\cite}

\def\Ref#1{(\ref{#1})}

\def\ft#1#2{{\textstyle {\frac{#1}{#2}} }}
\def\8{\infty}
\def\p{\partial}

\def\tr{{\rm tr \,}}
\def\ra{\rightarrow}
\def\i{{\rm i}}
\def\ad{{\rm ad}}
\def\16{N\!=\! 16}
\def\216{d\!=\!2 , N\!=\!16}
\def\316{d\!=\!3 , N\!=\!16}
\def\48{d\!=\!4 , N\!=\!8}

\def\cVh{\widehat{\cV}}
\def\Lh{\widehat{\L}}
\def\Qh{\widehat{Q}}
\def\Ph{\widehat{P}}

\def\GG{\mathbf{G}}
\def\GH{\mathbf{H}}
\def\gg{\mathfrak{g}}
\def\gh{\mathfrak{h}}
\def\gk{\mathfrak{k}}

\def\Ut{\widetilde{U}}
\def\Lt{\widetilde{L}}
\newcommand{\sr}{\gamma}

\makeatletter
\@addtoreset{equation}{section}
\makeatother

\begin{document}
\begin{flushright}
AEI-061\\hep-th/9804152
\end{flushright}

\vspace*{0.8cm}
\noindent
\rule{\linewidth}{0.6mm}
\vspace*{0.6cm}
\begin{center}
\mathversion{bold}
{\bf\LARGE  Integrability and Canonical Structure of
  $d\!=\!2, N\!=\!16$ Supergravity}\bigskip\\
\mathversion{normal}

{\bf\large H. Nicolai and H. Samtleben \medskip\\ }
{\large Max-Planck-Institut f\"ur Gravitationsphysik,\\
Albert-Einstein-Institut,\\
Schlaatzweg 1, D-14473 Potsdam, Germany}\smallskip\\ {\small E-mail:
nicolai@aei-potsdam.mpg.de, henning@aei-potsdam.mpg.de\medskip} 
\end{center}
\renewcommand{\thefootnote}{\arabic{footnote}}
\setcounter{footnote}{0}
\vspace*{0.6cm}
\noindent
\rule{\linewidth}{0.6mm}
\bigskip
\medskip
\begin{abstract}
  The canonical formulation of $\216$ supergravity is presented.  We
  work out the supersymmetry generators (including all higher order
  spinor terms) and the $\16$ superconformal constraint algebra. We
  then describe the construction of the conserved non-local charges
  associated with the affine $\9$ symmetry of the classical
  equations of motion.  These charges are shown to commute weakly with
  the supersymmetry constraints, and hence with all other constraints.
  Under commutation, they close into a quadratic algebra of Yangian
  type, which is formally the same as that of the bosonic theory. The
  Lie-Poisson action of $\9$ on the classical solutions is exhibited
  explicitly. Further implications of our results are discussed.
\end{abstract}
\newpage

\section{Introduction}
Maximal $\216$ supergravity is the most symmetric of all known field 
theories in two dimensions, and therefore of special interest in many 
ways. It is classically integrable in the sense that its equations of 
motion can be obtained from a linear system \ci{Nico87,NicWar89,Nico94}.
As first argued in \ci{Juli81} on the basis of a general analysis of the 
hidden symmetries arising in the dimensional reduction of $d\!=\!11$ 
supergravity to lower dimensions \ci{CrJuSc78,CreJul79}, the space of 
associated classical solutions admits an $\9$ symmetry generalizing 
the Geroch group of general relativity.\footnote{Following standard
usage, we will designate by $\9$ the relevant non-compact version 
of the affine Lie group $E_9$. However, most of our results will 
concern the associated Lie algebra, which we denote by $\eeee$.}
Its high degree of symmetry,
the emergence of a maximally extended superconformal structure and 
the natural appearance of exceptional groups of $E$-type indicate that 
this theory is destined to play a prominent role in the search for 
a non-perturbative and unified theory of quantum gravity encompassing 
superstring theory and $d\!=\!11$ supergravity.
 
In this paper we present the canonical formulation of $\216$
supergravity and analyze its symmetry structure within the canonical
framework. After a brief description of the Lagrangian and equations
of motion, we set up the canonical formalism and derive the complete
expressions for the $\16$ constraint generators of local
supersymmetry.  These results allow us in particular to complete the
proof that the linear system of \ci{Nico87,NicWar89} -- which is at most
quadratic in the fermionic fields -- generates all the necessary
higher order fermionic terms in the equations of motion. I.e.\ the
integrable structure of the model extends through all fermionic
orders.  We then proceed to work out the $\16$ superconformal
constraint algebra, which is not of the standard type.

In the second part of the paper, we analyze the integrable structure
of the model on the basis of an infinite set of conserved charges and
their algebra. These non-local charges are determined via the
transition matrices of the linear system. They are shown to commute
weakly with the supersymmetry constraints, and hence the full gauge
algebra; therefore they yield an infinite set of observables.  We
examine the canonical algebra that is generated by the non-local
charges, exploiting the fact that the final result of this calculation
is unambiguous -- unlike the corresponding calculation for flat space
integrable models of this type, which is plagued by irresolvable
ambiguities (see e.g.  \ci{VeEiMa84}). It is quite remarkable that the
Yangian algebra which one obtains turns out to be the same as that of
the purely bosonic model. This enables us to take over the analysis of
\ci{KorSam98} entirely.

The connection of our results with previous ones on the affine $\9$
symmetry of the classical solutions is finally established by defining
the Lie-Poisson action of the non-local charges on the physical
fields. This coincides with the known symmetry action on the fields
and their associated dual potentials. The main advantage of the
canonical realization of the affine symmetry is that in this way
we gain complete control over the deviations of the symmetry action 
from a symplectic action. This issue is of particular importance
when one studies the quantum mechanical realization of the symmetry. 
We believe that our results open new and promising perspectives for 
the exact quantization of $\216$ supergravity, as it is known at least 
in principle how to quantize Lie-Poisson actions \ci{BabBer92,Lu93}. 
In particular, they confirm the relevance of Yangian-type deformations 
of $\eeee$ for the classification of physical states in the quantum theory.

Our results could also be relevant in connection with recent
developments in non-perturbative string theory. There are fascinating
topics for future research in this direction, such as the search for
possible analogs of D-branes, which would require reconciling the
action of the Geroch group with boundary conditions other than
asymptotically flat ones, or the investigation of the possible
relevance of the symmetry structures found here for duality symmetries
in non-perturbative string theory.

\mathversion{bold}
\section{$N\!=\!16$ supergravity in two dimensions}
\mathversion{normal}

The Lagrangian and equations of motion of $\216$ supergravity
are most conveniently derived by dimensional reduction of $\16$ 
supergravity in three dimensions \ci{MarSch83} as described in 
\ci{Nico87,NicWar89,Nico94}. Let us first recall its field content. 
In the gravitational sector, we have the zweibein $e_\mu^{\;\a}$ and 
the dilaton $\r$, both of which originate from the dreibein 
of $\316$ supergravity. The Kaluza-Klein-Maxwell field $A_\mu$,
which is also part of the dreibein, is conventionally set to zero.
It does not carry propagating degrees of freedom, but may have non-trivial
holonomies on a topologically non-trivial worldsheet, and could give 
rise to a cosmological constant in two dimensions. Although we 
neglect such effects here, this field cannot be completely
ignored because its elimination gives rise to extra quartic spinor 
terms which do contribute to our complete expression for 
supersymmetry constraint below.

The dilaton $\r$ satisfies a free field equation (in the
gravitational background provided by $e_\mu^{\;\a}$), and
this permits us to introduce its dual ``axion'' $\tilde\r$ 
\be
\p_\mu \r + \e_{\mu\nu} \p^\nu \rt = 0\;.
\ee
The partner of the dreibein, the $d\!=\!3$ gravitino, gives rise 
to a gravitino in two dimensions and a ``dilatino'' according to
the decomposition $\psi^I_a=(\psi_\a^I, \psi_2^I)$ (in flat indices), 
both of which transform as the ${\bf 16}_v$ representation of $\SO$.

In addition to these non-propagating fields, there are 128 physical
scalar fields and 128 physical fermions transforming in the left and
right handed spinor representation of $\SO$, labeled by indices $A$
and $\dot A$, respectively. The chiral components associated with the
128 propagating fermionic degrees of freedom are designated by
$\chi^{\dot A}_\pm$ (see Appendix A.1 for our spinor conventions). 
The scalar sector is governed by a non-linear $\EE/\SO$ $\s$-model.
I.e.\ the scalar fields are described by a matrix $\cV(x)\in\E/\SO$ 
representing all 128 propagating bosonic degrees of freedom. 

For the Lagrangian and the equations of motion we need the
decomposition 
\be
\cV^{-1}\p_\m\cV = \ft12 Q^{IJ}_\m X^{IJ} + P^A_\m Y^A\;, \la{VdV}
\ee
where $X^{IJ}$ and $Y^A$ are the algebra generators of
$\mathfrak{e}_8$, see Appendix \ref{Ae8}.

The composite gauge field $Q_\mu^{IJ}$ serves to define the
$\SO$ covariant derivatives by 
\ba
D_\m \psi^I &=& \p_\m \psi^I + Q_\m^{IJ}\psi^J \;,\la{cov}\\
D_\m \chi^{\dot{A}} &=& \p_\m \chi^{\dot{A}} 
      +\ft14Q_\m^{IJ} \G^{IJ}_{\dot{A}\dot{B}}\chi^{\dot{B}} \;,\non
D_\m P_\n^A &=& \p_\m P_\n^A +
          \ft14 Q_\m^{IJ} \G^{IJ}_{AB} P_\n^B \;,\nn
\ea
and the field strength
\be
Q_{\mu\nu}^{IJ} := \p_\m Q_\n^{IJ}- \p_\nu Q_\mu^{IJ} 
         + Q_\m^{IK} Q_\n^{KJ} - Q_\m^{JK} Q_\n^{KI}\;.
\ee
{}From \Ref{VdV} one reads off the integrability relations
(valid in any dimension)
\be
D_{[\mu} P_{\nu]}^A = 0\;, \qquad  
Q_{\mu\nu}^{IJ} + \ft12 \G^{IJ}_{AB} P_\mu^A P_\nu^B = 0 \;.
\la{int}\ee

The Lagrangian of $\216$ supergravity can now be directly obtained 
from the one given in \ci{MarSch83} and reads
\ba
\cL &=& 
 -\ft14 \rho e R + 
    \rho \e^{\mu\nu} \overline{\psi}^I_2 D_\mu \psi_\nu^I - 
    \ft12 \i \rho e \overline{\chi}^{\dot A} \g^\mu D_\mu \chi^{\dot A}
    + \ft14 \r e P^{\mu A} P^A_\mu \la{L2}\\                 
  &&{}
       - \ft12 \rho e  \overline{\chi}^{\dot A} \g^\nu \g^\mu \psi^I_\nu
               \G^I_{A\dot A} P^A_\mu  
       - \ft12 \i \rho e  \overline{\chi}^{\dot A} \g^3 \g^\mu \psi^I_2
               \G^I_{A\dot A} P^A_\mu \;. \nn
\ea
up to higher order fermionic terms. The associated action is manifestly
invariant under general coordinate transformations in two dimensions,
as well as local $\SO$ transformations
\ba
\d_\o\, Q_\pm^{IJ} &=& D_\pm \o^{IJ} ~=~
\p_\m \o^{IJ}+ Q_\m^{IK} \o^{KJ} - Q_\m^{JK} \o^{KI}
\;,\la{so16}\\
\d_\o\, P_\pm^A &=& 
 \ft14\G^{IJ}_{AB}\o^{IJ}P_\pm^B \;,\non
\d_\o\, \psi^I &=& \o^{IJ}\psi^J \;,\non
\d_\o\, \chi^{\dot{A}} &=& \ft14 \G^{IJ}_{\dot{A}\dot{B}}\o^{IJ} 
\chi^{\dot{B}} \;,\nn
\ea
with the $\SO$-parameter $\o^{IJ}(x)=-\o^{JI}(x)$.

For our further considerations we employ the superconformal gauge
\be\la{sconf}
e_\mu^{\;\a} = \l  \d_\mu^\a\;, \qquad 
\psi_\mu^I=\i\g_\mu \psi^I\;.
\ee
We will also make use of the fields 
\be
\s := \log \l \qquad \widehat\s := \s - \ft12\log (\p_+\r \p_-\r)
\ee
The redefined field $\widehat\s$ transforms as a genuine scalar under
conformal diffeomorphisms, whereas $\l$ itself is a density.
The gauge choice \Ref{sconf} must be accompanied by the following 
rescaling of the fermion fields: 
\be
\chi^{\dot A} \ra \l^{\ft12} \chi^{\dot A}\;, \qquad
\psi^I_\mu \ra \l^{\ft12} \psi_\mu^I\;, \qquad
\psi^I_2 \ra \l^{\ft12} \psi_2^I \;.
\ee
Then the conformal factor $\l$ disappears almost entirely from 
the Lagrangian, except for the Einstein 
term which in the conformal gauge becomes
\be
-\ft14 \rho e R = -\ft12\p_\mu \rho \p^\mu \s\;.
\ee
As a consequence the theory would be conformally invariant if it 
were not for the remaining $\s$ dependence of this term. Similarly,
the superpartner $\psi^I$ of $\s$ does not completely decouple  
in the gauge \Ref{sconf}, as it would in a superconformally 
invariant theory. Still, we can from now on drop the distinction 
between flat and curved indices $\mu,\dots$ and $\a,\dots$ 
We remark that there is no problem of principle in keeping the 
dependence on topological degrees of freedom which are 
eliminated by \Ref{sconf} (i.e. the moduli and supermoduli on the 
worldsheet); the requisite formalism has been set up in 
\ci{Nico94}, building on earlier results in \ci{Dick89,Dick92}.

We next list the equations of motion in the  superconformal gauge
\Ref{sconf}. 
Utilizing one-component spinors (see Appendix \ref{Aspin}), the 
fermionic equations of motion read
\ba
D_\pm(\r^{\frac12}\chi_\mp^{\dot{A}}) &=& 
\mp\ft12\r^{\frac12} \psi^{I}_{2\,\mp}\G^{I}_{A\dot{A}}P_\pm^A 
\;,\la{feqm}\\
D_\pm\psi^I_\mp &=& -\ft12 \chi_\mp^{\dot{A}}
\G^{I}_{A\dot{A}}P_\pm^A \;,\non
D_\pm(\r\psi^{I}_{2\,\mp}) &=& 0 \;,\nn
\ea
modulo cubic spinor terms. 
The equations of motion for the physical scalar fields are
\ba
D_+(\rho P_-^A) + D_- (\rho P_+^A) &=& 
2\i\G^I_{A\dot{A}} D_-(\r\psi^I_{2\,+}\chi^{\dot{A}}_+)
-2\i\G^I_{A\dot{A}} D_+(\r\psi^I_{2\,-}\chi^{\dot{A}}_-)
\la{beqm}\\[2pt]
&&{}+2\i\r\G^{IJ}_{AB}P^B_-\psi^I_{2\,+}\psi^J_+
-2\i\r\G^{IJ}_{AB}P^B_+\psi^I_{2\,-}\psi^J_-   \non[2pt]
&&{}+\ft14\i\r\G^{IJ}_{AB} P_-^B \,
\chi^{\dot{A}}_+ \G^{IJ}_{\dot{A}\dot{B}}\chi^{\dot{B}}_+ 
+\ft14\i\r\G^{IJ}_{AB} P_+^B \, 
\chi^{\dot{A}}_- \G^{IJ}_{\dot{A}\dot{B}}\chi^{\dot{B}}_-\;,\non[5pt]
\p_+\p_- \widehat \s ~=~ \p_+\p_-\s &=& - \ft12P^A_+P^A_-
-\i \left(\chi^{\dot{A}}_+D_-\chi^{\dot{A}}_+ +
            \chi^{\dot{A}}_-D_+\chi^{\dot{A}}_- \right)\;,\nn
\ea
modulo quartic spinor terms. 

The equations listed so far originate from the variation of those fields
which survive the superconformal gauge fixing. There are, however,
two more equations that follow by variation of the traceless 
modes of the metric and the gravitino, both of which are put to zero 
in the superconformal gauge. These are the constraints 
\ba
T_{\pm\pm} &=& \ft12\r P_\pm^AP_\pm^A - \p_\pm\r\,\p_\pm\hat{\s}
\mp\i\r P_\pm^A \G^I_{A\dot{B}}\psi_{2\,\pm}^I\chi_\pm^{\dot{B}} 
-\i\r\chi_\pm^{\dot{A}} D_\pm\chi_\pm^{\dot{A}}\la{T}\\
&&{}\mp\i \psi_{\pm}^ID_\pm(\r\psi_{2\,\pm}^I)
\mp\i \r\psi_{2\,\pm}^ID_\pm\psi_{\pm}^I
\hspace*{10em}~~\approx~ 0\;,\non[5pt]   \la{T1}
S_\pm^I &=&
\pm D_1(\r\psi^I_{2\,\pm})-\r\p_\pm\s\,\psi^I_{2\,\pm} 
\mp\r\chi^{\dot{A}}_\pm\G^I_{A\dot{A}}P_\pm^A 
\pm \p_\pm\r\,\psi^I_\pm\hspace*{2.7em}~\approx~ 0\;, \la{S1}
\ea
which will be seen to generate conformal and superconformal
transformations, respectively. For this reason, 
they enjoy a somewhat different status from the previous 
equations: the equations setting them to zero must 
be interpreted as weak equalities in the sense of Dirac (indicated 
by the symbol ``$\approx$''). Like in superconformal field theories
the derivatives of these constraints vanish strongly:
\be
\p_\mp T_{\pm\pm} ~=~ D_\mp S_\pm^I ~=~ 0\;.
\ee
While we do not include the cubic spinor terms in the above equations,
we will give the complete terms  below in the canonical expression for
the supersymmetry constraint. 

The superconformal gauge \Ref{sconf} is preserved by local 
supersymmetry transformations with chiral parameters
$\e^I_\pm$ obeying
\be
D_\pm\e^I_\mp=0\;,
\ee
again modulo cubic spinor terms. Up to such terms, the
supersymmetry variations are given by
\be\la{susy}
\begin{array}{rclrcl}
\cV^{-1}\d_\pm \cV &=& 
\mp 2\i\e_\pm^I\chi^{\dot{A}}_\pm\G^{I}_{\dot{A}A} Y^A \;, & 
\hspace*{3.5em}\d_\pm \chi^{\dot{A}}_\pm &=& \mp
\e^I_\pm\G^{I}_{A\dot{A}} P_\pm^{A} \;,\\[6pt]
\d_\pm \r &=& 2\i\r \e^I_\pm\psi_{2\,\pm}^I \;,  &
\d_\pm \psi^{I}_{2\,\pm} &=& \r^{-1}\p_\pm\r \e^I_\pm \;,\\[6pt]
\d_\pm \s &=& \mp2\i\e^I_\pm\psi^I_\pm \;,  &
\d_\pm \psi_{\pm} &=& \mp
\left( D_\pm^{I}\e_\pm^I + \p_\pm\s\e_\pm^I\right)\;.
\end{array}
\ee
The superalgebra generated by these transformations can be regarded as
an $N=16$ superconformal algebra, but it is distinguished from the
standard superconformal algebras (which stop at $N=4$) by the fact
that it is a {\em soft} algebra (i.e.\ it has field dependent
structure ``constants''). This is one of the reasons why, despite the
evident similarities with superconformal field theories, $\16$
supergravity belongs to a different class of models. There is
no immediate analog of the left-right (holomorphic) factorization 
of conformal field theory because $\16$ is not a free (or even 
``quasi-free'') theory. \footnote{However, there is an analog of
holomorphic factorization within the framework of isomonodromic 
solutions \ci{KorNic96}.} We will recognize this feature again 
when analyzing the constraint algebra. 

We conclude this section with some comments on the physical
interpretation of the fields $\r$ and $\rt$, which play a special
role. As is well known, one can invoke the residual conformal
invariance of \Ref{sconf} to identify these fields with the worldsheet
coordinates at least locally. This gauge, which is the precise analog
of the light-cone gauge in string theory, is complemented on the
fermionic side by the elimination of the dilatino $\psi^I_2$ from the
equations of motion by means of the residual superconformal
transformations \Ref{susy} \ci{Nico87}. For stationary axisymmetric
solutions of Einstein's equations (Euclidean signature of the
worldsheet), one traditionally takes $\r \geq 0$ as the radial
variable and $\rt$ as the coordinate along the symmetry axis
\ci{KSHM80}. For the Minkowskian signature considered here, more and
physically distinct choices are possible. Depending on whether the
vector $\p_\mu \r$ is spacelike or timelike, we can either identify
$\r$ with the radial and $\rt$ with the time coordinate
(Einstein-Rosen gravitational waves \ci{EinRos37}), or $\r$ with the
time and $\rt$ with the space coordinate (in which case there is a
``big bang'' singularity at $\r =0$ \ci{Gowd74}). However, one can
also envisage more general situations with alternating signatures, as
well as worldsheets of non-trivial topology.

{}From a ``stringy'' perspective, on the other hand, $\r$ and $\rt$
should be treated as (quantum) fields living on the worldsheet, whose
vacuum expectation values would be associated with coupling constants
of the theory. The M\"obius subgroup of the hidden Witt-Virasoro
symmetry of the theory \ci{HouLee87,Mais88,JulNic96} is then 
analogous to the strong-weak coupling duality in string theory.

\section{Canonical brackets}
The derivation of the of the canonical brackets from the
action \Ref{L2} is straightforward, except perhaps for the
technical complication that the presence of second class 
constraints necessitates a Dirac procedure.

In the gravitational sector we have the canonical momenta
\ben
2\ps = \p_0 \rho\;, \qquad  2 \pr = \p_0 \s\;.
\een
In order not to overburden the notation, we will set
\be
\p_\pm\s\equiv \pi_\r \pm \ft12 \p_1 \s\;, \qquad
\p_\pm\r\equiv \pi_\s \pm \ft12 \p_1 \r\;,
\ee
in the formulas below. Furthermore, we will be exclusively 
concerned with equal time brackets at a fixed but arbitrary 
time $t\equiv x^0$, and will therefore not explicitly indicate 
the full coordinate dependence, but only spell out the dependence 
on the space coordinates $x\equiv x^1, y\equiv y^1$.

The equal time brackets for the dilaton $\r$ and the 
conformal factor $\s$ are given by
\be
\left\{\r(x)\,,\,\pr(y)\right\} =
\left\{\s(x)\,,\,\ps(y)\right\} = \d(x\!-\!y)\;.\la{PBb1}
\ee
Duality implies that
\be\la{PBb1a}
\left\{\rt(x)\,,\,\p_1\s(y)\right\} = 2\,\d(x\!-\!y)\;,
\ee
and
\be
\left\{\p_\pm\r(x)\,,\p_\pm\s(y)\right\} = \pm\d'(x\!-\!y) \qquad , \qquad
\left\{\p_\pm\r(x)\,,\p_\mp\s(y)\right\} = 0
\ee
where derivatives on the $\d$ function are always understood to
act on the first argument. Later, we will also introduce a spectral 
parameter $\sr$ (see \Ref{sr}) depending on both $\rho$ and $\rt$. 
In the canonical framework, this spectral parameter becomes a
canonical variable, and thus an operator upon quantization.

To obtain the canonical bosonic Poisson brackets in the $\s$-model
sector, we introduce conjugate momenta $\Pi$ to the canonical variables
$Q_1$ and $P_1$, 
\ben
\Pi^{IJ} \equiv \frac{\d S}{\d (\p_0Q_1^{IJ})} \;,\qquad
\Pi^{A} \equiv \frac{\d S}{\d (\p_0P_1^{A})} \;.
\een
with
\be\la{PBc}
\left\{Q_1^{IJ}(x),\Pi^{KL}(y)\right\} ~= ~
\d^{IJ}_{KL}\,\d(x\!-\!y)\;,\quad
\left\{P_1^A(x),\Pi^B(y)\right\} ~=~ \d^{AB}\,\d(x\!-\!y)\;.
\ee
Taking into account the zero curvature condition \Ref{int}, the
Lagrangian \Ref{L2} yields 
\ba
\p_1\Pi + \left[\cV^{-1}\p_1\cV\,,\,\Pi\right] &=&
\ft1{2}\,\r P_0^A Y^A +\ft{1}{2}\,\i\r\,
\G^I_{A\dot{A}}\overline{\chi}^{\dot{A}}\g_1\psi_{2}^I \,Y^A
  + \i\r\,\overline{\psi}^I\g_1\psi_2^J\,X^{IJ}\non
&&{}-\ft{1}{8}\,\i\r\,\overline{\chi}^{\dot{A}}\g_0
\chi^{\dot{B}}\,\G^{IJ}_{\dot{A}\dot{B}}X^{IJ}\;,\nn
\ea
where $\Pi$ is given by
\ben
\Pi \equiv - \Pi^{IJ}X^{IJ}+ \Pi^AY^A\;.
\een
Solving the above relations for $P_0^A$, we arrive at
\be\la{rel}
P_0^A ~=~
\frac{2}{\r}\left(D_1\Pi^A 
+ \ft14 \Pi^{IJ}\G^{IJ}_{AB} P_1^B\right) 
+2\i\G^I_{A\dot{A}}\psi^I_{2\,+}\chi^{\dot{A}}_+\,Y^A
-2\i\G^I_{A\dot{A}}\psi^I_{2\,-}\chi^{\dot{A}}_-\,Y^A \;.
\ee
Furthermore, we deduce the (first class) $\SO$ constraint:
\ba
\Phi^{IJ} &=& D_1\Pi^{IJ} 
+\ft14 \G^{IJ}_{AB}P^A_1\Pi^B \la{con}\\
&&{} -2\i\r\left(\psi^{[I}_+\psi_{2\,+}^{J]}
-\psi^{[I}_-\psi_{2\,-}^{J]}\right) 
-\ft{1}4\i\r\G^{IJ}_{\dot{A}\dot{B}}
\left(\chi_+^{\dot{A}}\chi_+^{\dot{B}}+
\chi_-^{\dot{A}}\chi_-^{\dot{B}}\right)~\approx~ 0 \;. 
\nn
\ea
which generates the gauge transformations \Ref{so16}.
{}From \Ref{rel} we obtain the bosonic
brackets in the $\s$-model sector:
\ba
\left\{P_\pm^A(x)\,,\cV(y)\right\} &=&
\frac1{\r}\cV(x)Y^A\:\d(x\!-\!y)\;,\la{PBb2}\\
\left\{P_\pm^A(x)\,,Q_1^{IJ}(y)\right\} &=& 
-\frac1{4\r}\;\G^{IJ}_{AB}P_1^B \:\d(x\!-\!y)\;,
\non 
\left\{P_\pm^A(x)\,,P_\pm^B(y)\right\}&=& 
\mp\frac1{4\r}\;\G^{IJ}_{AB}Q_1^{IJ} \:\d(x\!-\!y)
    \mp
    \frac12\left({\frac1{\r(x)}+\frac1{\r(y)}}\right)
\;\d^{AB}\:\d'(x\!-\!y)\non
&&{}-\frac{\i}{2\r}\;\G^{IJ}_{AB}
\left(2\psi^I_+\psi_{2\,+}^{J}-2\psi^I_-\psi_{2\,-}^{J}
+\psi^I_{2\,+}\psi_{2\,+}^{J}+\psi^I_{2\,-}\psi_{2\,-}^{J}\right)
\:\d(x\!-\!y)\non
&&{}-\frac{\i}{8\r}\;\G^{IJ}_{AB}\G^{IJ}_{\dot{A}\dot{B}}
\left(\chi_+^{\dot{A}}\chi_+^{\dot{B}}+
\chi_-^{\dot{A}}\chi_-^{\dot{B}}\right)\:\d(x\!-\!y)\non
&&{}+\frac1{4\r^2}\;\G^{IJ}_{AB}\Phi^{IJ}\:\d(x\!-\!y)\;,\non 
\left\{P_\pm^A(x)\,,P_\mp^B(y)\right\}&=& 
\pm\frac1{2\r^2}\:\p_1\r\:\d^{AB}\,\d(x\!-\!y)
{}+\frac1{4\r^2}\;\G^{IJ}_{AB}\Phi^{IJ}\:\d(x\!-\!y)\non 
&&{}-\frac{\i}{2\r}\;\G^{IJ}_{AB}
\left(2\psi^I_+\psi_{2\,+}^{J}-2\psi^I_-\psi_{2\,-}^{J}
+\psi^I_{2\,+}\psi_{2\,+}^{J}+\psi^I_{2\,-}\psi_{2\,-}^{J}\right)
\:\d(x\!-\!y)\non
&&{}-\frac{\i}{8\r}\;\G^{IJ}_{AB}\G^{IJ}_{\dot{A}\dot{B}}
\left(\chi_+^{\dot{A}}\chi_+^{\dot{B}}+
\chi_-^{\dot{A}}\chi_-^{\dot{B}}\right)
\:\d(x\!-\!y)\;,\non
&&\non
\left\{P_\pm^A(x)\,,\p_\pm\s(y)\right\} &=& 
-\frac1{2\r}\left(\,
  P_0^A+\i\G^I_{A\dot{B}}
\left(\psi_{2\,+}^I\chi^{\dot{B}}_+-\psi_{2\,-}^I\chi^{\dot{B}}_-\right)
\right) 
\d(x\!-\!y) \;,\la{PBb3} \\
\left\{P_\pm^A(x)\,,\p_\mp\s(y)\right\} &=& 
-\frac1{2\r}\left(\,
  P_0^A+\i\G^I_{A\dot{B}}
\left(\psi_{2\,+}^I\chi^{\dot{B}}_+-\psi_{2\,-}^I\chi^{\dot{B}}_-\right)
\right) \d(x\!-\!y) \;.\nn
\ea
In the fermionic sector we find the following Dirac brackets
\ba
\left\{\chi^{\dot{A}}_\pm(x)\,,
\chi^{\dot{B}}_\pm(y)\right\} &=&
-\frac{\i}{2\r}\;\d^{\dot{A}\dot{B}} \,\d(x\!-\!y) \;,\la{PBf}\\
\left\{\psi^I_\pm(x)\,,
\psi_{2\,\pm}^J(y)\right\} &=&\mp\frac{\i}{2\r}\;\d^{IJ} 
\:\d(x\!-\!y) \;.\nn
\ea
Owing to the explicit appearance of the bosonic fields in the fermionic
second class constraints there are also non-vanishing mixed brackets
\ba
\left\{\pr (x) \,,\chi_\pm^{\dot{A}}(y)\right\} &=&
\frac1{2\r}\:\chi_\pm^{\dot{A}}\,\d(x\!-\!y) \;,\la{PBbf}\\
\left\{\pr (x)\,,\psi_{2\,\pm}^I(y)\right\} &=&
\frac1{\r}\:\psi_{2\,\pm}^I\,\d(x\!-\!y) \;,\nn
\ea
while the form of $P_0^A$ in \Ref{rel} gives rise to
\ba
\left\{P_0^A(x)\,,\chi_\pm^{\dot{B}}(y)\right\} &=&
\pm\frac1{\r}\G^I_{A\dot{B}}\psi_{2\,\pm}^I\,\d(x\!-\!y)\;,\la{PBbf2}\\
\left\{P_0^A(x)\,,\psi_{\pm}^I(y)\right\} &=&
-\frac1{\r}\G^I_{A\dot{B}}\chi_{\pm}^{\dot{B}}\,\d(x\!-\!y)\;.\nn
\ea
Alternatively, the necessity of \Ref{PBbf} can be inferred from
the presence of $\rho$ on the r.h.s.\ of \Ref{PBf}).

The above brackets are slightly simplified when written in terms
of canonical variables with vanishing mixed brackets. For this 
purpose, we introduce
\be
\widetilde P_0^A ~:=
 \r\,P_0^A - 2\i\r\,\psi_{2+}^I \chi_+^{\dot A} \G^I_{A\dot A} 
  + 2 \i\r\,\psi_{2-}^I \chi_-^{\dot A} \G^I_{A\dot A}\;,
\ee
together with
\ben
\widetilde P^A_\pm := \ft12( \widetilde P_0^A \pm P_1^A) \;.
\een
These are the variables which commute with all the fermions 
and with $\p_\pm\s$. Moreover, we notice that $\psi^I$ and
the rescaled fermions $\r\psi_2^I$ and $\r^{\frac12}\chi^{\dot A}$ 
commute with $\pr$, and hence $\p_\pm\s$ as well.

\section{Constraint superalgebra}
In this section we establish the constraint superalgebra underlying
the superconformal transformations \Ref{susy} which remain after
imposing the superconformal gauge \Ref{sconf}. They are shown to close
into a superconformal algebra which in addition contains the conformal
transformations generated by \Ref{T} and the $\SO$ gauge
transformations \Ref{con}.

Our most important result in this section is the expression 
for the supersymmetry constraint generators $S_\pm^I$ \Ref{Spm} 
below, which is complete and includes all cubic fermionic terms.  
This is the crucial operator because all other constraints
can be completely determined from the commutator of two 
supersymmetry generators (in principle we could thus compute
the quartic spinorial contributions to $T_{\pm\pm}$, but the explicit
expressions are not very illuminating). In the next section
we will make use of these results and demonstrate that the integrals 
of motion associated with the affine $\9$ symmetry of the 
equations of motion weakly commute with the supersymmetry
constraint. This calculation provides a stringent consistency check 
on the correctness of the cubic spinor terms in $S_\pm^I$.

As discussed above, the energy momentum (Virasoro) constraints
$T_{\pm\pm}$ descend from \Ref{L2} before going into the
superconformal gauge \Ref{sconf}. For the determination of 
canonical brackets and the constraint algebra it is, 
however, necessary to express $T_{\pm\pm}$ and the other
constraints entirely in terms of canonical variables. In other words,
all time derivatives implicit in the derivatives $\p_\pm$ must be
converted into momenta and spatial derivatives of the canonical
variables by means of their equations of motion.  More specifically,
the fermionic equations of motion \Ref{feqm} can be invoked to derive
\ba
D_\pm(\r^{\frac12}\chi^{\dot{A}}_\pm) &=& 
\pm D_1(\r^{\frac12}\chi^{\dot{A}}_\pm) 
\pm\ft12\r^{\frac12}\psi_{2\,\pm}^I\G^I_{A\dot{A}}P_\mp^A \;,\non
D_\pm \psi^I_\pm &=& \pm D_1 \psi^I_\pm
-\ft12\chi^{\dot{A}}_\pm\G^I_{A\dot{A}}P_\mp^A \;,\non
D_\pm(\r\psi_{2\,\pm}^I) &=&
\pm D_1(\r\psi^I_{2\,\pm})\;,\nn
\ea
where the l.h.s. is to be replaced by the expression on the r.h.s.
before computing any bracket. The resulting form of the constraint
will be referred to as the ``canonical form''.  Nevertheless, writing
the constraints in ``non-canonical'' form \Ref{T}, \Ref{S1} has the
advantage that the conformal covariance properties become
manifest. Consequently, we will use both forms of the constraints
according to convenience.

The canonical form of the energy momentum constraint is given by
\ba
T_{\pm\pm} &=& \ft12\r P_\pm^AP_\pm^A - \p_\pm\r\,\p_\pm\s 
\pm\ft12\p_1\p_\pm\r 
\mp\i\r P_1^A \G^I_{A\dot{B}}\psi_{2\,\pm}^I\chi_\pm^{\dot{B}} \\
&&{}\mp\i\r\chi_\pm^{\dot{A}} D_1\chi_\pm^{\dot{A}}
 - \i \psi_{\pm}^ID_1(\r\psi_{2\,\pm}^I)
- \i \r\psi_{2\,\pm}^ID_1(\psi_{\pm}^I)\;,\nn   \la{Tpm}
\ea
again up to quartic fermionic terms. The supersymmetry constraint 
reads in canonical form
\ba
S_\pm^I &=&
\pm D_1(\r\psi^I_{2\,\pm})-\r\p_\pm\s\,\psi^I_{2\,\pm} 
\mp\r\chi^{\dot{A}}_\pm\G^I_{A\dot{A}}P_\pm^A 
\pm \p_\pm\r\,\psi^I_\pm\la{Spm}\\
&&{}\mp\i\r\psi_{\pm}^J\;\chi_\pm\G^{IJ}\chi_\pm
-\ft{1}2 \i\:\r\psi_{2\,\pm}^J\G^{IJ}_{\dot{A}\dot{B}}
\left(\chi^{\dot{A}}_\pm\chi^{\dot{B}}_\pm-
\chi_\mp^{\dot{A}}\chi_\mp^{\dot{B}}\right)\non
&&{}+2\i\r\psi_{\pm}^I\psi_{\pm}^J\psi_{2\,\pm}^J
\pm2\i\r\psi_{\mp}^I\psi_{2\,\pm}^J\psi_{2\,\mp}^J
\mp2\i\r\psi_{2\,\mp}^I\psi_{\mp}^J\psi_{2\,\pm}^J
-2\i\r\psi_{2\,\mp}^I\psi_{2\,\pm}^J\psi_{2\,\mp}^J\;.\nn
\ea
In contradistinction to the formula for $T_{\pm\pm}$ we have
given the complete expression including all cubic fermionic
terms. These extra terms have been derived by requiring closure of the
superalgebra, see \Ref{PBSS1}, \Ref{PBSS2} below, and it is important
here that this method does fix the higher order terms uniquely. 
Alternatively, they could have been determined from the full
equations of motion, but with more effort, since this would have
required analyzing the quartic terms resulting from the elimination of
the Kaluza-Klein-Maxwell vector as well. 

Local conformal $N\!=\!16$ supersymmetry variations are generated
according to
\be 
\d_\pm \varphi = 2\i  \int dx\, \e_\pm^I(x) \left\{S_\pm^I (x) \,,\,
                     \varphi  \right\}\;,
\ee
Readers are invited to check that the resulting variations coincide 
with the ones stated in \Ref{susy} to the relevant order. 

The local supersymmetry generators $S^I_\pm$ satisfy the constraint
algebra  
\ba
\left\{S_\pm^I(x)\,,S_\pm^J(y)\right\} &=&
-\d^{IJ}\Big(
\i T_{\pm\pm}\mp2\psi_\pm^KS_\pm^K 
-\ft14\chi_\pm^{\dot{A}}\chi_\pm^{\dot{B}}\G^{KL}_{\dot{A}\dot{B}}
\Phi^{KL}\Big)\:\d(x\!-\!y)\la{PBSS1}\\
&&{}\mp\Big(\psi_\pm^IS_\pm^J 
+ \psi_\pm^JS_\pm^I\Big)\:\d(x\!-\!y)\non  
&&{}+\ft12\chi_\pm^{\dot{A}}\chi_\pm^{\dot{B}}
\Big(\G^{IK}_{\dot{A}\dot{B}}\Phi^{KJ}+
\G^{JK}_{\dot{A}\dot{B}}\Phi^{KI}\Big)
\:\d(x\!-\!y)\;,\nn
\ea
\ba
\left\{S_+^I(x)\,,S_-^J(y)\right\} &=&
-\d^{IJ}\Big(\psi_{2\,+}^KS_-^K +  \psi_{2\,-}^KS_+^K 
\Big)\:\d(x\!-\!y)\la{PBSS2}\\
&&{}+\Big(\psi_{2\,-}^IS_+^J + \psi_{2\,+}^JS_-^I
\Big)\:\d(x\!-\!y)\non
&&{}+\ft14\chi_+^{\dot{A}}\chi_-^{\dot{B}}\,
\G^I_{\dot{A}A}\G^{KL}_{AB}\G^{J}_{B\dot{B}}\,\Phi^{KL}\:
\d(x\!-\!y)\;.\nn
\ea
This is the complete result, valid in all fermionic orders. As
mentioned above, the constraint superalgebra closes into the
energy momentum constraints $T_{\pm\pm}$ and the $\SO$ constraints
$\Phi^{IJ}$. The closure of the algebra is, of course, guaranteed 
by general consistency arguments, but the canonical algebra has 
so far not been exhibited explicitly for this theory. It is conceivable 
that it can be further simplified by making field dependent 
redefinitions of the constraints.

Conformal coordinate transformations with parameters 
$\xi^\pm = \xi^\pm (x^\pm)$ are generated by  
\be\la{conformal}
\d_{\xi^\pm}\, \varphi  ~= ~
\int dx \,\xi^\pm (x)\left\{ T_{\pm\pm}(x)\,,\,\varphi \right\}~
 =~ -h_\varphi^{\pm}\,\p_\pm\xi^\pm\,\varphi 
+\xi^\pm \,D_\pm \varphi\;, 
\ee 
where $h_\varphi^{\pm}$ denotes the conformal dimensions of the field
$\varphi$. 
This formula illustrates the interplay between the canonical and the
covariant framework. Canonically, the gauge parameters $\xi^\pm$ are
defined as functions of and integrated over the spatial coordinate
$x$. Upon using the equations of motion for $\varphi$ and restoring
the time dependence of $\xi^\pm$ according to $\p_\pm\xi^\mp\!=\!0$,
the r.h.s.\ of \Ref{conformal} takes a conformally covariant form.  
Thus, $T_{\pm\pm}$ indeed generates translations along the
$x^\pm$ coordinates modulo local $\SO$ transformation with field
dependent parameter $Q_1^{IJ}$ (here we have tacitly adopted the 
Coulomb gauge $Q_0^{IJ}\!=\!0$). The algebra \Ref{PBSS1} permits 
us to calculate \Ref{conformal} and thus the equations of motion in
all fermionic orders by means of the super-Jacobi identities.  

The constraints \Ref{con} generate the $\SO$ transformations
\Ref{so16} via   
\be
\d_\o\, \varphi \equiv \int \!dx\; 
  \o^{IJ}(x)\left\{ \Phi^{IJ}(x)\,, \varphi \right\}\;.
\ee
and satisfy the $\SO$ algebra:
\be
\left\{\Phi^{IJ}(x)\,,\Phi^{KL}(y)\right\} ~=~ 
\Big(\d^{JK}\Phi^{IL} - \d^{IK}\Phi^{JL} 
+ \d^{IL}\Phi^{JK} - \d^{JL}\Phi^{IK}\Big)\,\d(x\!-\!y) \;.
\ee

The remaining commutation relations of the superconformal algebra
are listed below
\ba
\left\{T_{\pm\pm}(x)\,,T_{\pm\pm}(y)\right\} &=& 
\mp\left(T_{\pm\pm}(x)+T_{\pm\pm}(y)\right)\,\d'(x\!-\!y)\;, 
\la{Vir}\\[2pt]
\left\{T_{\pm\pm}(x)\,,T_{\mp\mp}(y)\right\} &=& 
\ft14\,\G^{IJ}_{AB}P_+^AP_-^B\Phi^{IJ} \,\d(x\!-\!y)\;,\non[2pt]
\left\{T_{\pm\pm}(x)\,,S_\pm^I(y)\right\} &=& \mp\ft32
S_\pm^I(y)\,\d'(x\!-\!y) + D_\pm S_\pm^I \,\d(x\!-\!y)\non[1pt]
&&{}\mp\ft14\,\big(\G^{KL} \G^{I}\big)_{A\dot{A}}
P_\pm^A\chi^{\dot{A}}_\pm\Phi^{KL} \,\d(x\!-\!y)\;,\non[2pt]
\left\{T_{\pm\pm}(x)\,,S_\mp^I(y)\right\} &=& 
\pm\ft14\,\big(\G^{KL} \G^{I}\big)_{A\dot{A}}
P_\pm^A\chi^{\dot{A}}_\mp\Phi^{KL} \,\d(x\!-\!y)\;,\non[2pt]
\left\{\Phi^{IJ}(x)\,,S_\pm^K(y)\right\} &=& 
\ft12\left(\d^{IK}S^J_\pm-\d^{JK}S^I_\pm\,\right)\,\d(x\!-\!y) 
\;,\non[2pt]
\left\{\Phi^{IJ}(x)\,,T_{\pm\pm}(y)\right\} &=& 0 \;.\nn
\ea
We have not worked out the higher order fermionic contributions
even though in principle all of them can be determined
straightforwardly (though tediously) from \Ref{Spm} and the
super-Jacobi identities.

The constraint superalgebra \Ref{PBSS1}, \Ref{PBSS2}, \Ref{Vir} is a
superconformal extension of the Virasoro algebra \Ref{Vir} with
$N\!=\!16$ supersymmetry. Its existence does not contradict the
well-known absence of the standard superconformal algebras with
$N\!>\!4$ \ci{RamSch76}, because in comparison with the algebras which
have been studied in superconformal field theory, it exhibits some
unusual features. Let us briefly comment on the differences.

First of all, unlike the usual superconformal algebras, the present
model does not completely factorize into two chiral halves: this
is already evident from the equations of motion, but also reflected
by the fact that the chiral supercharges $S_+$ and $S_-$ 
in \Ref{PBSS2} do not commute with one another. 

Secondly, as already pointed out before, the brackets \Ref{PBSS1} 
and \Ref{PBSS2} do not close into a linear algebra. Rather, on the r.h.s\ 
the constraints $S_\pm^I$ appear with coefficients that explicitly
depend on the fermionic fields $\psi^I$ and $\psi_2^I$. This was of
course to be expected in view of the general result that the algebras
arising in supergravity are usually ``soft'' gauge algebras
\ci{Nieu81,Sohn83}. The important new feature here is that
we can nevertheless give a complete canonical characterization.

Finally, we stress the conspicuous absence of internal chiral current
algebras which appear in the standard extended superconformal algebras
\ci{RamSch76}. Namely, a linear superconformal chiral algebra with
$N\!\leq\!4$ supercharges requires chiral internal bosonic currents
multiplying $\d'(x\!-\!y)$ on the r.h.s.\ of \Ref{PBSS1}. This fact
can be immediately deduced from the super-Jacobi identities involving
$\{S^I,\{S^J,S^K\}\}$, whose $\d'$ terms only cancel with the
contribution from the additional current. By contrast, we here have
only one ``vectorlike'' $\SO$ current $\Phi^{IJ}$; moreover, in the 
algebra this current appears only in second order in the 
fermions and multiplies the $\d$-function rather than its derivative. 
The terms required for consistency of the super-Jacobi 
identity now originate from the additional contributions due to 
the field dependent structure constants 
on the r.h.s. of \Ref{PBSS1}. Another distinctive feature is the 
invariance of the generators $T_{\pm\pm}$ under $\Phi^{IJ}$; 
the $\SO$ constraint generator thus carries zero conformal weight, 
unlike the chiral currents of standard superconformal field theory.

A further extension of our results which we postpone to later
investigations would involve relaxing the superconformal gauge
and thus entail a canonical treatment (and quantization) of the 
topological degrees of freedom as well.

\section{Conserved charges}
Our main purpose in this section is to investigate the infinitely many
conserved charges associated with the classical $\9$ symmetry of
the equations of motion, their algebra and the infinite dimensional
symmetries they generate. Furthermore, we demonstrate the
supersymmetry invariance of these charges by showing that they weakly
commute with the full supersymmetry constraints, and hence with all
other constraints.

\subsection{Linear system}
As shown in \ci{Nico87,NicWar89,Nico94} the supergravity equations of
motion can be obtained as the compatibility condition of a linear 
system (or Lax pair) for an $E_8$-valued matrix $\cVh(x;\sr)$. 
Here $\sr$ is a spectral parameter which depends explicitly on
the coordinates via the dilaton and the axion fields, as we will
shortly explain. The field dependence of $\sr$ is in marked contrast
to the constancy of the spectral parameter appearing in flat space 
integrable systems such as nonlinear $\s$-models and their rigidly 
supersymmetric extensions. 

In the version of \ci{NicWar89}, the linear system takes the form 
\be\la{ls}
\cVh^{-1} \p_\pm\cVh(\sr) ~=~L_\pm(\sr)~\equiv~
\ft12\Qh^{IJ}_\pm(\sr)X^{IJ}+\Ph^{A}_\pm(\sr)Y^A \;,
\ee
with the connection coefficients
\ba
\Qh^{IJ}_\pm(\sr) &=& Q^{IJ}_\pm 
-\frac{2\i\sr}{(1\!\pm\!\sr)^2}\; 
\left(8\psi_{2\,\pm}^{[I}\psi^{J]}_\pm \pm \G^{IJ}_{\dot{A}\dot{B}} 
\chi_\pm^{\dot{A}}\chi_\pm^{\dot{B}} 
\right) 
-\frac{32\i\sr^2}{(1\!\pm\!\sr)^4}\;
\psi_{2\,\pm}^{I}\psi_{2\,\pm}^{J} \;,\non
\Ph^{A}_\pm(\sr) &=& \frac{1\!\mp\!\sr}{1\!\pm\!\sr}\;P_\pm^A
+\frac{4\i\sr(1\!\mp\!\sr)}{(1\!\pm\!\sr)^3}\;
\G^{I}_{A\dot{B}}\psi_{2\,\pm}^{I}\chi^{\dot{B}}_\pm \;.\nn
\ea
Despite the occurrence of cubic and quartic spinor terms in the 
fermionic and bosonic equations of motion, the linear system \Ref{ls} 
does not receive any higher order corrections but is at most 
quadratic in the fermionic fields. In other words, {\em the linear
system generates all required higher order fermionic terms by itself}. 
So far, this had only been demonstrated in the ``super-Weyl gauge'' 
where only the $(\bar\chi\chi)^2$ terms had to be checked \ci{Nico87}. 
The more general result, which also includes the higher order
contributions involving the gravitinos and dilatinos, is a  
consequence of the result \Ref{US} in the next section.

The spectral parameter $\sr$ is subject to the differential equations
\ci{Mais78,BelZak78,BreMai87}
\be\la{diff}
\sr^{-1}\p_\pm \sr ~=~ \frac{1\mp\sr}{1\pm \sr} \,\r^{-1}\p_\pm \r\;.
\ee
At a practical level, this field dependence is due to the 
appearance of the nonconstant dilaton field $\r$ in the 
equations of motion listed in section 2. At a more fundamental 
level it is linked to the presence of a hidden Witt-Virasoro 
symmetry of the theory and the fact that (a reformulation
of) equation \Ref{diff} may be interpreted as a linear system 
for the dilaton itself \ci{JulNic96}. An important
consequence of the dependence of $\sr$ on $\r$ and $\rt$ 
is that in this way the spectral parameter becomes a canonical 
variable of its own, having non-vanishing canonical brackets
with the conformal factor.

The solution of \Ref{diff} depends not only on the fields $\rho$ 
and $\rt$, but also on an integration constant $w$, that is
sometimes called ``constant spectral parameter'' (because it is
the parameter relevant for the $\eeee$ current algebra). It is 
given by
\be\la{sr}
\sr(\r,\rt;w) ~=~
\frac1{\r}\left(w+\rt-\sqrt{(w+\rt)^2-\r^2}\;\right)
\quad\Longleftrightarrow\quad
w~=~\ft12 \r \left(\sr + \frac1{\sr}\right) - \rt    \;. 
\ee
The function $\sr(\r,\rt;w)$ lives on the two-sheeted covering 
of the complex $w$ plane with a field-dependent branch cut
connecting the points $w_\pm=-\rt\pm\r$ on the real $w$-axis. 
This cut disappears in the limits $\r\ra 0$ or $\rt\ra\8$,
whereas it extends over the whole real axis for $\r\ra\8$,
in which case the two sheets disconnect from one another.
For the later treatment, we will be particularly interested 
in the critical points where the function $\sr(\r,\rt;w)$ becomes
independent of $w$. This happens at
\begin{equation}\la{limits} 
\sr(\r\!\rightarrow\!0) ~\rightarrow~
\left\{\begin{array}{l}
    \!\!0\\
    \!\!\8
                  \end{array}\right. \;,\quad
\sr(\r\!\rightarrow\!\8) ~\rightarrow~ \left\{\begin{array}{r} 
                               \i\\
                               -\i
                  \end{array}\right. \;,\quad
\sr(\rt\!\rightarrow\!\pm\8) ~\rightarrow~ \left\{\begin{array}{l} 
                               \!\!0\\
                               \!\!\8 
                  \end{array}\right.\;,
\end{equation}
where the two values correspond to the two sheets of the
covering. The transition between the two sheets is performed by
$\sr\mapsto\frac1{\sr}$. 

The algebra involution $\t$ from \Ref{tau} can be extended to an
involution $\t^\8$ which acts on $E_8$-valued functions of the
spectral parameter $\sr$ by combining the action on $E_8$ with a
transition between the two sheets of $\sr$ \ci{Juli83,BreMai87}
\be\la{genin}
\t^\infty\Big(U(\sr)\Big)\equiv
\t\left(U\left(\frac1{\sr}\right)\right)\;.
\ee
This involution leaves the the linear system \Ref{ls} invariant.

\subsection{Non-local conserved charges}
The non-local charges are obtained from the transition 
matrices\footnote{In this equation, we write $x\equiv x^1, y\equiv y^1$,
etc. for the space coordinates as before.} 
\ba
U(t,x,y;w) &\equiv& \cP \exp
\int_x^y\!dz\,L_1\Big(t,z;\sr(t,z;w)\Big)\la{trans}\\
&\equiv& \cVh^{-1}\big(t,x;\sr(t,x;w)\big)
     \cVh\big(t,y;\sr(t,y;w)\big)\quad\in E_8\;,
\ea
associated with the linear system \Ref{ls}. For $w\!\notin\!\R$ (so as
to avoid any ambiguities caused by the branch cut on the real line)
these transition matrices are defined uniquely and like $\sr$ live on
a two-sheeted covering of the complex $w$-plane. Like the connection
of the linear system they are invariant under the generalized
involution $\t^\8$ \Ref{genin}.  Unless specified otherwise, we will
consider the sheet underlying the unit disc of the complex
$\sr$-plane.

We further define the following objects
\be\la{Ut}
\Ut(t,x,y;w) ~\equiv~ \cV(t,x)U(t,x,y;w)\cV^{-1}(t,y)\;, 
\quad\mbox{for}\;\;w\notin\R\;,
\ee
and the monodromy matrix \ci{BreMai87}
\ba\la{M}
\cM(t,z;w) &\equiv& \lim_{\e\ra +0} 
\bigg(\cV(z)U(t,z,y;w\!+\!\i\e)\;
\t\!\left(U(t,y,z;w\!-\!\i\e)\cV(z)^{-1}\right)\bigg)\;,\\
&&\mbox{for}\;\;w\in\R\;,\quad|w+\rt(t,z)|>|\r(t,z)|\;,
\mbox{and}\;\;|w+\rt(t,y)|<|\r(t,y)|\;.\nn
\ea
Unless $\r$ and $\rt$ are constant fields (in which case the solution
becomes trivial anyway) one may always find points $z$, $y$ which
satisfy the last two conditions. Mathematically, these conditions
mean, that with coordinates $(t,z)$ the spectral parameter $\sr$ is
single valued in a neighborhood of $w$, whereas with coordinates
$(t,y)$ the parameter $w$ lies on the branch cut of \Ref{sr}, implying
that $w\!+\!\i\e$ and $w\!-\!\i\e$ tend to $\sr(y,w)$ and
$\sr^{-1}(y,w)$, respectively, see the figure. The condition on $y$
thus guarantees that the definition of $\cM$ indeed does not depend on
$y$; the condition on $z$ then basically ensures the nontriviality of
$\cM$.

\begin{figure}[htbp]
  \begin{center}
    \leavevmode \input{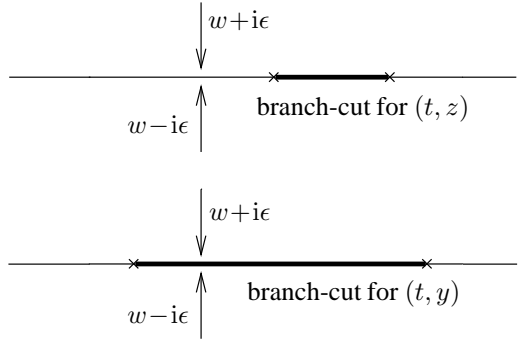}
  \end{center}
  \caption{Illustrating the definition of the monodromy matrix $\cM$
    in the $w$-plane} 
  \label{planes}
\end{figure}

According to their definition, $\Ut(t,x,y;w)$ and $\cM(t,z;w)$ have
the following time dependence:
\ba
\p_t \widetilde{U}(t,x,y;w) &=& 
-\widetilde{L}_0(t,x,\sr(t,x,w))\,\widetilde{U}+ 
\widetilde{U}\,\widetilde{L}_0(t,y,\sr(t,y,w)) \;,\la{timeU}\\
\p_t \cM(t,z;w) &=& 
-\widetilde{L}_0(t,z,\sr(t,z,w))\,\cM+
\cM\:\t\!\left(\widetilde{L}_0(t,z,\sr(t,z,w))\right)\;,\la{timeM}
\ea
with
\ben
\widetilde{L}_0 ~=~ 
\cV L_0\cV^{-1} -\p_0\cV\cV^{-1} \;.
\een
Thus, the modified transition matrix $\Ut(t,x_0,y_0;w)$ becomes time
independent, and hence an integral of motion, if it connects two
points $x_0$ and $y_0$ at both of which $\Lt_0$ vanishes.  Similarly,
the monodromy matrix $\cM(z_0;w)$ turns into an integral of motion if
$\Lt_0$ vanishes at $z_0$. There are two ways in which this can
happen. Either the physical fields vanish at these points (as they
would for instance at spatial infinity), or otherwise the spectral
parameter $\sr$ vanishes (cf. \Ref{limits}) with the physical fields
remaining regular. For instance, both situations are realized for
cylindrical gravitational waves, where $\r\!=\!0$ corresponds to the
origin, and $\r\!=\!\8$ to spatial infinity. What we would like to
emphasize, however, is that the present framework allows for more
general possibilities: depending on the behavior of the dilaton field
(i.e. the zeroes and poles of the functions $\r$ and $\rt$) there
might even exist several conserved charges.\footnote{From the Kaluza
  Klein point of view, the values $\r=0$ and $\8$, respectively,
  correspond to internal manifolds of zero or infinite size.}

The matrix $\cM(z_0;w)$ is of special interest. As a function of real
$w$ it is single-valued, real and satisfies 
\be\la{Msym}
\cM(w) = \t\left(\cM^{-1}(w)\right)\;.
\ee
We may further introduce its Riemann-Hilbert decomposition 
\be\la{Upm}
\cM(w) \equiv U_+(w)\,\t\!\left(U_-^{-1}(w)\right)\;,
\ee
into $E_8$-valued functions $U_\pm(w)$ which are holomorphic in the
upper and the lower half of the complex $w$-plane, respectively.  
For instance, with the gravitational wave boundary conditions mentioned 
above, the functions $U_\pm(w)$ are related to the modified
transition matrices \Ref{Ut} \ci{KorSam98} by
\be
U_{\!\pm}(w)~=~\Ut(0,\8;w) \quad\mbox{for 
$\;\;\Im w{{}_{\textstyle>}\atop {}^{\textstyle<}}0$}\;.
\ee
It is important here that the analytic continuation of $U_+$
into the lower half plane does not coincide with $U_-$, 
and vice versa. Rather they are related by
\ben
U_{\!+}(w)~=~\overline{U_{\!-}(\bar{w})}\;.
\een
The monodromy matrix $\cM(w)$ in this case 
yields the values of the original physical fields on the 
symmetry axis $\r\!=\!0$ for $\rt =w\in\R$.

We note two basic differences with the flat space integrable models.
First, the coordinate dependence of the spectral parameter $\sr$ has
given rise to the definition of the monodromy matrix $\cM$ \Ref{M}
which has no analog in flat space. Secondly, \Ref{timeU} shows that
for spatially periodic boundary conditions, the eigenvalues of the
transition matrices are not necessarily integrals of motion.  The
reason is that periodicity of the physical fields is not enough to
ensure periodicity of the dual potentials because even if we choose
the the dilaton $\r$ to be a periodic function, the spectral parameter
$\sr$ will not be periodic due to the non-periodicity of the dual
axion field $\rt$. It remains an open problem to reconcile periodicity
with the existence of infinite dimensional duality symmetries.

\bigskip
\subsection{Supersymmetry of non-local charges}
As we have already pointed out, in a theory with {\em local}
supersymmetry, time independence is not quite enough to distinguish
reasonable observables (in the sense of Dirac). In addition, these
must weakly commute with the full gauge algebra
\Ref{PBSS1}--\Ref{Vir}.  The particular constraint associated with
time translations (alias the Wheeler-DeWitt operator 
in the quantized theory) is just one part of this gauge algebra, 
and follows from the commutation of two supersymmetry constraints.

In this section we will show that the integrals of motion identified
above are indeed invariant under the full gauge algebra.
We first note that the modified transition matrices \Ref{Ut} and the
monodromy matrix \Ref{M} for
arbitrary values of $x$, $y$ and $z$ are invariant under the $\SO$
gauge transformations \Ref{so16} generated by $\Phi^{IJ}$ from
\Ref{con}: 
\be
\left\{\Phi^{IJ}(u)\,,\Ut(x,y;w)\right\}~=~ 
\left\{\Phi^{IJ}(u)\,,\cM(z;w)\right\}~=~0\;.
\ee

The main result in this section is the behavior of the transition
matrices \Ref{trans} under supersymmetry transformations: 
\be
\Big\{U(x,y;w),S^I_\pm(z)\Big\} ~~=~~
\frac{4\sr\,\th(x,z,y)}{\r(1\pm\sr)^2}\;
U(x,z;w)X^{IJ}S^J_\pm(z)U(z,y;w)\:\la{US}
\ee
\vspace*{-1.3em}
\ba
\qquad\qquad\qquad\quad
&\pm&\frac{\sr\,\th(x,z,y)}{\r(1-\sr^2)}\;\chi_\pm^{\dot{B}}\,
\big(\G^{I} \G^{JK}\big)_{\dot{B} A}\:\Phi^{JK}U(x,z;w)Y^AU(z,y;w)\:\non
&\pm&\frac{1\mp\sr}{1\pm\sr}\;
\chi_\pm^{\dot{B}}\,\G^{I}_{A\dot{B}}
\left(U(x,y;w)Y^A\d(z\!-\!y)-Y^AU(x,y;w)\d(z\!-\!x)\right)\non
&\mp&\frac{4\sr}{(1\pm\sr)^2}\;\psi_{2\,\pm}^J
\left(U(x,y;w)X^{IJ}\d(z\!-\!y)-X^{IJ}U(x,y;w)\d(z\!-\!x)\right)
\;,\nn
\ea
with
\be\la{theta}
\th(x,z,y):=\left\{\begin{array}{ll} 
                                1&\mbox{for~} x<z<y\;\\
                                0&\mbox{else}\;\;(x\not=y\not=z)
                  \end{array}\right.\;.
\ee 

This result is valid in all orders of fermions, i.e.\ including all
the cubic fermionic terms from \Ref{Spm}.
For the modified transition matrices $\Ut$ and the monodromy matrix
$\cM$ it implies:
\ba
\left\{\Ut(x,y;w),S^I_\pm(z)\right\} &\approx&
\frac{2\sr}{1\pm\sr}\;\chi_\pm^{\dot{B}}\,\G^{I}_{A\dot{B}}\,
\left(\cV Y^A\cV^{-1}\right)\Ut(x,y;w)\,\d(z\!-\!x)\la{UtS}\\
&&{}
-\frac{2\sr}{1\pm\sr}\;\chi_\pm^{\dot{B}}\,\G^{I}_{A\dot{B}}\,
\Ut(x,y;w)\left(\cV Y^A\cV^{-1}\right)\,\d(z\!-\!y)\non
&&{}\pm\frac{4\sr}{(1\pm\sr)^2}\;\psi_{2\,\pm}^J\,
\left(\cV X^{IJ}\cV^{-1}\right)\Ut(x,y;w)\,\d(z\!-\!x)\non
&&{}\mp\frac{4\sr}{(1\pm\sr)^2}\;\psi_{2\,\pm}^J\,
\Ut(x,y;w)\left(\cV X^{IJ}\cV^{-1}\right)\,\d(z\!-\!y)\;,\nn
\ea
and
\ba
\Big\{\cM(z;w),S^I_\pm(x)\Big\} &\approx&
\frac{2\sr}{1\pm\sr}\;\chi_\pm^{\dot{B}}\,\G^{I}_{A\dot{B}}
\Big(\!\left(\cV Y^A\cV^{-1}\right)\cM -
\cM\,\t\!\left(\cV Y^A\cV^{-1}\right) \!\Big)\,\d(x\!-\!z) \non
&&{}\pm\frac{4\sr}{(1\pm\sr)^2}\,\psi_{2\,\pm}^J
\Big(\!\left(\cV X^{IJ}\cV^{-1}\right)\cM +
\cM\,\t\!\left(\cV X^{IJ}\cV^{-1}\right) \!\Big)\,\d(x\!-\!z)\;. \nn
\ea
The r.h.s.\ of these equations vanishes under the very same conditions
that have been discussed in \Ref{timeU}, \Ref{timeM} for the vanishing
of $\Lt_0$.  This shows that the integrals of motion obtained in the
previous section are indeed superconformally invariant. Due to the
form of the constraint superalgebra \Ref{PBSS1}, the brackets of these
charges with $T_{\pm\pm}$ then also vanish weakly. They are thus
invariant under the full gauge algebra.

In particular, this implies our previous claim that the linear system
\Ref{ls} does not receive any quartic corrections but captures the
full content of the theory. Since by supersymmetry transformations 
\Ref{susy} any solution can be fixed to obey the super-Weyl gauge, 
the invariance of the linear system under supersymmetry shows 
that indeed no quartic corrections arise in the general case.

The rest of this section is devoted to an outline of the proof of
\Ref{US}. It is obtained from the general formula 
\ben
\left\{U(x,y;v)\,,S^I_\pm(z')\right\}
~=~\int_x^y dz\: U(x,z;v)\left\{
L_1\big(z,\sr(z;v)\big)\,,S^I_\pm(z')\right\} U(z,y;v) \;,
\een
or equivalently
\begin{equation}\la{TSg}
U(z'\!,x,v)\left\{U(x,y;v)\,,S^I_\pm(z')\right\}U(y,z'\!;v)~=~
\end{equation}
\ben
\hspace*{12em}\int_x^y dz\: U(z'\!,z,v)\left\{
L_1\big(z,\sr(z;v)\big)\,,S^I_\pm(z')\right\} U(z,z'\!,v) \;.
\een

It is straightforward although lengthy to evaluate \Ref{TSg} using the
form of the supersymmetry generator \Ref{Spm} and the fundamental
Poisson brackets \Ref{PBb1}--\Ref{PBbf2}. Up to the higher order terms
in the fermions this result had already been given in \ci{Nico94}.
Thus it remains to check the cubic fermionic terms. As we can show,
all the extra cubic terms cancel with the exception of those required
to complete the supersymmetry generators in \Ref{US} to the full
expressions given in \Ref{Spm}.

There are altogether four different sources yielding cubic fermionic
terms. First such terms come from the brackets involving cubic terms
in the supersymmetry generators $S^I_\pm$, second from bilinear
fermionic terms in the Poisson brackets \Ref{PBb2} between $P_0$ and
$P_0$.  Third, they arise from the Poisson brackets involving
$\p_\pm\s$ in $S^I_\pm$ and at last, cubic terms enter when partial
integration of the $\d'$ terms in \Ref{TSg} leads to the appearance of
the connection $L_1$ again.

To give an idea of the calculation we display the cancellation of the
cubic terms proportional to $\psi_{2\,\pm}\psi_{2\,\pm}\chi_\pm$ in
\Ref{TSg}. 
According to \Ref{PBf} and \Ref{PBbf2} we have
\ben
\left\{L_1(\sr)\,,\chi_\pm^{\dot{A}}\right\} ~=~ 
\frac{\sr}{\r(1\!\pm\!\sr)^2}\,
\G^{IJ}_{\dot{A}\dot{B}}\chi_\pm^{\dot{B}}X^{IJ}\d(z\!-\!z')
-\frac{8\i\sr^2}{\r(1\!\pm\!\sr)^2(1\!-\!\sr^2)}\,
\G^K_{A\dot{A}}\psi_{2\,\pm}^K Y^A\d(z\!-\!z')\;,
\een
such that the cubic term $\psi_{2\,\pm}\chi_\pm\chi^{\dot{B}}_\pm$
from \Ref{Spm} gives the contribution
\be\la{s1} 
\left\{L_1(\sr)\:,-\ft12 \i\:\r\psi_{2\,\pm}^K
\chi_\pm\G^{IK}\chi_\pm\right\}~\ra~
\frac{-8\i\sr^2}{\r(1\!\pm\!\sr)^2(1\!-\!\sr^2)}\;
\big(\G^{IK} \G^{L}\big)_{\dot{A} A}
\psi_{2\,\pm}^K\psi_{2\,\pm}^L\chi^{\dot{A}}Y^A \;,
\ee
to the r.h.s\ of \Ref{TSg}. Next, there comes a contribution from the
bracket between $P_0$ in $L_1(\sr)$ and the $\r\chi_\pm P_\pm$ part
of the supersymmetry constraint \Ref{Spm}, which is due to the
quadratic fermionic terms in \Ref{PBb2} and reads  
\be\la{s2}
\left\{L_1(\sr)\:,\mp\r\chi^{\dot{A}}_\pm\G^I_{A\dot{A}}P_\pm^A
\right\} ~\ra~
\frac{-8\sr^2}{\r(1\!\pm\!\sr)^2(1\!-\!\sr^2)}\;
\big(\G^{KL}\G^{I}\big)_{A\dot{A}}
\psi_{2\,\pm}^K \psi_{2\,\pm}^L\chi^{\dot{A}}Y^A \;.
\ee
Making use of \Ref{Gamma} the two terms \Ref{s1} and \Ref{s2} sum up to 
\be\la{s12}
\frac{8\i\sr^2}{(1\!\pm\!\sr)^2(1\!-\!\sr^2)}\;\G^K_{A\dot{A}}\;
\psi_{2\,\pm}^I\psi_{2\,\pm}^K\chi^{\dot{A}}Y^A\;.
\ee

Several further relevant terms arise from the Poisson brackets
involving the $\r\psi_{2\,\pm}\p_\pm\s$ term in \Ref{Spm}. Namely,
$\{L_1(\sr),\pr \}$ gives rise to several bilinear fermionic terms
due to the brackets \Ref{PBb3}, \Ref{PBbf} and eventually also due to 
\ben
\{\sr(z)\,,\p_\pm\s(z')\} ~=~ 
\frac{\sr(1\!\mp\!\sr)}{\r(1\!\pm\!\sr)}\;\d(z\!-\!z')\;.
\een
Altogether they sum up to
\be\la{s3}
\left\{L_1(\sr)\,,-\r\p_\pm\s\,\psi^I_{2\,\pm} \right\}~\ra~ 
\frac{16\i\sr^2(1\mp4\sr+\sr^2)}{1\!\pm\!\sr)^4(1\!-\!\sr^2)}\;
\G^{K}_{A\dot{A}}
\psi_{2\,\pm}^I\psi_{2\,\pm}^K \chi^{\dot{A}}Y^A \;.
\ee
Finally, the integrand of \Ref{TSg} has terms proportional to
$\d'(z\!-\!z')$ due to 
\ben
\left\{L_1(t)\:,\mp\r\chi^{\dot{A}}_\pm\G^I_{A\dot{A}}P_\pm^A
\right\}~\ra~ 
\pm\frac{1\!\mp\!\sr}{1\!\pm\!\sr}\;
\G^{I}_{A\dot{A}}\chi_\pm^{\dot{A}}Y^A \,\d'(z\!-\!z') \;, 
\een
and
\ben
\left\{L_1(t)\:,\pm\r\p_1\psi_{2\,\pm}^I\right\}~\ra~ 
\pm \frac{4\sr}{(1\!\pm\!\sr)^2}\;\psi_{2\,\pm}^KX^{KI} 
\,\d'(z\!-\!z')\;.
\een
Upon partial integration in \Ref{TSg} and using \Ref{ls} they give
rise to 
\be\la{s4}
\mp\frac{1\!\mp\!\sr}{2(1\!\pm\!\sr)}\;\Qh^{KL}_\pm(\sr)
\G^{I}_{A\dot{A}}\chi_\pm^{\dot{A}}\:\left[X^{KL},Y^A\right]~\ra~
\frac{8\i\sr^2(1\!\mp\!\sr)}{(1\!\pm\!\sr)^5}\;
\big(\G^{KL}\G^I\big)_{A\dot{A}}
\psi_{2\,\pm}^K \psi_{2\,\pm}^L \chi^{\dot{A}}Y^A 
\ee
and
\be\la{s5}
\mp\frac{2\sr}{(1\!\pm\!\sr)^2}\;
\Ph^{A}_\pm(\sr)\psi_{2\,\pm}^K\:\left[Y^A,X^{KI}\right] ~\ra~
\frac{8\i\sr^2(1\!\mp\!\sr)}{(1\!\pm\!\sr)^5}\;
\big(\G^{KI}\G^L\big)_{A\dot{A}}
\psi_{2\,\pm}^K \psi_{2\,\pm}^L \chi^{\dot{A}}Y^A \;.
\ee
The sum of \Ref{s4} and \Ref{s5} then yields (again with some
$\G$-matrix algebra \Ref{Gamma})
\be\la{s45}
\frac{-24\i\sr^2(1\!\mp\!\sr)}{(1\!\pm\!\sr)^5}\;\G^K_{A\dot{A}}
\psi_{2\,\pm}^I\psi_{2\,\pm}^K \chi^{\dot{A}}Y^A\;.
\ee
Adding the different terms \Ref{s12}, \Ref{s3} and \Ref{s45} finally
leads to
\be
\frac{8\i\sr^2}{(1\!\pm\!\sr)^2(1\!-\!\sr^2)}~+~
\frac{16\i\sr^2(1\mp4\sr+\sr^2)}{(1\!\pm\!\sr)^4(1\!-\!\sr^2)}~-~
\frac{24\i\sr^2(1\!\mp\!\sr)}{(1\!\pm\!\sr)^5} ~~=~~ 0\;.
\ee
We see, how the terms of the type $\psi_{2\,\pm}\psi_{2\,\pm}\chi_\pm$
from all the different sources eventually cancel. In a similar way all
the unwanted cubic fermionic terms in \Ref{TSg} can be shown to drop
out.

We have thus found an infinite number of observables in the 
sense of Dirac.  We note that a similar transformation behavior 
has been observed in the supersymmetric extension of the 
nonlinear $\s$-model \ci{CorZac79,CurZac80,AbdFor86,SalZac97}. 
There, with suitable boundary conditions the bosonic non-local charges 
are invariant under global supersymmetry. In our model, invariance 
under the local supersymmetry is an indispensable condition for 
meaningful observables, since supersymmetry appears as a constraint.

\section{Algebra of charges and symmetries}
 
\subsection{Algebra of conserved charges}
We now calculate the Poisson algebra of the conserved charges that 
we have derived above. As it turns out, it is entirely sufficient 
to compute the brackets for the connection coefficients entering 
the linear system \Ref{ls}. Our key result is that the fermionic
contributions conspire in precisely such a way that the canonical
brackets \Ref{LL} below {\it are identical with the ones obtained
for the purely bosonic theory}! This implies that, as far as the
analysis of conserved non-local charges and their algebra
is concerned, we can take over the analysis of the bosonic case 
in \ci{KorSam98} practically without modification. However, the
realization of the algebra and hence the quantum spaces on which 
the algebra eventually acts will be very different in the two cases.

The starting point of the computation is the well-known formula 
\ba
\left\{\stackrel1{U}\!(x,y;v)\;,\,\stackrel2{U}\!(x',y';w)\right\}
&=&
\int_x^y\!dz\int_{x'}^{y'}\!dz'\enspace
\Big(\!\stackrel1{U}\!(x,z;v)\stackrel2{U}\!(x',z';w)\Big) \la{n}\\
&& \hspace{-2.2em}
\left\{\stackrel1{L_1}\!(z,\sr(z,v))\;,
\stackrel2{L_1}\!(z',\sr(z',w))\right\}
\Big(\!\stackrel1{U}\!(z,y;v)\stackrel2{U}\!(z',y';w)\Big)\;,\nn
\ea
which we have given in the tensor notation explained in Appendix
\ref{Aten}.  

Further evaluation requires the following canonical brackets
between the connection coefficients in the linear system \Ref{ls}:
\ba
\left\{\Qh^{IJ}_1(\sr_1),\Qh^{KL}_1(\sr_2)\right\} &=&
-\frac{4\sr_1\sr_2}{\r(\sr_1\!-\!\sr_2)(1\!-\!\sr_1\sr_2)}
\;\d(x\!-\!y)\;\times\\
&&{}\Bigg(\d^{JK}\!\left(\Qh_1^{IL}(\sr_1)\!-\!\Qh_1^{IL}(\sr_2)\right)-
\d^{IK}\!\left(\Qh_1^{JL}(\sr_1)\!-\!\Qh_1^{JL}(\sr_2)\right) \non
&&{}+\d^{IL}\!\left(\Qh_1^{JK}(\sr_1)\!-\!\Qh_1^{JK}(\sr_2)\right)-
\d^{JL}\!\left(\Qh_1^{IK}(\sr_1)\!-\!\Qh_1^{IK}(\sr_2)\right)\Bigg)\;,
\non[4pt]
\left\{\Qh^{IJ}_1(\sr_1),\Ph^A_1(\sr_2)\right\} &=&
\frac{2\sr_2^2(1\!-\!\sr_1^2)}
{\r(1\!-\!\sr_2^2)(\sr_1\!-\!\sr_2)(1\!-\!\sr_1\sr_2)}\;
\G^{IJ}_{AB}\Ph_1^B(\sr_1)\:\d(x\!-\!y)\non
&&{}-\frac{2\sr_1\sr_2}{\r(\sr_1\!-\!\sr_2)(1\!-\!\sr_1\sr_2)}\;
\G^{IJ}_{AB}\Ph_1^B(\sr_2)\:\d(x\!-\!y)\;,\non[4pt]
\left\{\Ph^A_1(\sr_1),\Ph^B_1(\sr_2)\right\} &=&
-\frac{(1\!-\!\sr_1^2)\sr_2^2}
{\r(1\!-\!\sr_2^2)(\sr_1\!-\!\sr_2)(1\!-\!\sr_1\sr_2)}\;
\G^{IJ}_{AB}\Qh_1^{IJ}(\sr_1)\:\d(x\!-\!y)\non
&&{}+\frac{(1\!-\!\sr_2^2)\sr_1^2}
{\r(1\!-\!\sr_1^2)(\sr_1\!-\!\sr_2)(1\!-\!\sr_1\sr_2)}\;
\G^{IJ}_{AB}\Qh_1^{IJ}(\sr_2)\:\d(x\!-\!y)\non
&&{}+\frac{4\d^{AB}}{(1\!-\!\sr_1^2)(1\!-\!\sr_2^2)}
\left(\frac{\sr_1(1\!+\!\sr_2^2)}{\r(x)}+
\frac{\sr_2(1\!+\!\sr_1^2)}{\r(y)}\right)
\:\d'(x\!-\!y)\non
&&{}+\frac{4\sr_1\sr_2}{\r^2(1\!-\!\sr_1^2)(1\!-\!\sr_2^2)}\;
\G^{IJ}_{AB}\,\Phi_{IJ}\:\d(x\!-\!y)\;.\nn
\ea
Here we have introduced the shorthand notation
$\sr_1\!\equiv\!\sr(x,v)$, $\sr_2\!\equiv\!\sr(y,w)\,$
(as usual we suppress the dependence on the time coordinate $t$ here).
It is convenient to combine these equations into a single one
by means of the index-free tensor notation introduced in Appendix
\ref{Aten}: 
\ba
\left\{\stackrel1{L}_1\!(\sr_1)\;,\;
\stackrel2{L}_1\!(\sr_2)\right\} &\approx& 
-\frac{4\,\O_\gk }{(1\!-\!\sr_1^2)(1\!-\!\sr_2^2)}
\left(\frac{\sr_1(1\!+\!\sr_2^2)}{\r(x)}+
\frac{\sr_2(1\!+\!\sr_1^2)}{\r(y)}\right)
\d'(x-y)\;,\la{LL}\\
&&{}-\frac{4\sr_1\sr_2}{\r(\sr_1-\sr_2)(1-\sr_1\sr_2)}
\left[\O_\so\;,\; \stackrel1{L}_1\!(\sr_1)
+\!\stackrel2{L}_1\!(\sr_2)\right]\d(x-y)\non
&&{}-\frac{4\sr_2^2(1-\sr_1^2)}
{\r(1-\sr_2^2)(\sr_1-\sr_2)(1-\sr_1\sr_2)}
\left[\O_\gk\;,\;\stackrel1{L}_1\!(\sr_1)\right]\d(x-y) \non
&&{}- \frac{4\sr_1^2(1-\sr_2^2)}
{\r(1-\sr_1^2)(\sr_1-\sr_2)(1-\sr_1\sr_2)}
\left[\O_\gk\;,\;\stackrel2{L}_1\!(\sr_2) \right]\d(x-y)\nn
\ea
where we have dropped the contribution containing $\Phi^{IJ}$ so that
the equality holds only on the constraint hypersurface.  These
brackets coincide with the ones of the purely bosonic theory
\ci{KorSam98}. This fortuitous circumstance enables us to take over
the result for the bracket of two transition matrices \Ref{Ut} from
\ci{KorSam98}. Namely, inserting the above relations into \Ref{n} and
taking into account the additional contributions of the type
\ben
\left\{\stackrel1{U}(x,y;w)\, , \,\stackrel2{\cV}(z) \right\}~=~
 \frac{2\sr(z;w)}{\r(z)(1-\sr^2(z;w))}\;\th(x,z,y)
  \stackrel1{U}(x,z;w) \stackrel2{\cV}(z)\O_\gk
  \stackrel1{U}(z,y;w)\;,
\een
finally leads to the following algebra:
\begin{equation} \la{PBU}
\stackrel1{\cV^{-1}}\!(x)\!\stackrel2{\cV^{-1}}\!(x')
\left\{\stackrel1{\Ut}\!(x,y;v)
\;,\,\stackrel2{\Ut}\!(x'\!,y'\!;w)\right\} 
\stackrel1{\cV}\!(y)\!\stackrel2{\cV}\!(y') 
\end{equation}
\vspace*{-1em}
\ba
&=&\hspace{1em}{}\frac2{v-w}\;\times\;\;\left\{\;\;
\th(x,x'\!,y)\;\Big(\stackrel1{U}\!(x,x';v)
\;\O_\so\;
\stackrel1{U}\!(x'\!,y;v)\!\stackrel2{U}\!(x'\!,y';w)\Big)\right.\non
&&\hspace{6em}
{}+\th(x'\!,x,y')\;
\Big(
\stackrel2{U}\!(x'\!,x;w)\;\;\O_\so\;
\stackrel1{U}\!(x,y;v)\!\stackrel2{U}\!(x,y';w)\Big)\non
&&\hspace{6em}{}-\th(x,y'\!,y)\;
\Big(\stackrel1{U}\!(x,y';v)
\stackrel2{U}\!(x'\!,y';w)\;\;\O_\so\;
\stackrel1{U}\!(y'\!,y;v)\Big)\non
&&\hspace{6em}{}\left.-\th(x'\!,y,y')\;
\Big(\stackrel1{U}\!(x,y;v)
\stackrel2{U}\!(x'\!,y;w)\;\;\O_\so\;
\stackrel2{U}\!(y,y';w)\Big)\;\right\}\non
&&{}+\frac{2\th(x,x'\!,y)}{v-w}\;f(x';w,v)\,
\Big(\!\stackrel1{U}\!(x,x';v)
\;\O_\gk
\stackrel1{U}\!(x'\!,y;v)\!\stackrel2{U}\!(x'\!,y';w)\!\Big)\non
&&{}+\frac{2\th(x'\!,x,y')}{v-w}\;f(x;v,w)\,
\Big(\!
\stackrel2{U}\!(x'\!,x;w)\;\O_\gk
\stackrel1{U}\!(x,y;v)\!\stackrel2{U}\!(x,y';w)\Big)\non
&&{}-\frac{2\th(x,y'\!,y)}{v-w}\;f(y';w,v)\,
\Big(\!\stackrel1{U}\!(x,y';v)\!
\stackrel2{U}\!(x'\!,y';w)\;\O_\gk
\stackrel1{U}\!(y'\!,y;v)\Big)\non
&&{}-\frac{2\th(x'\!,y,y')}{v-w}\;f(y;v,w)\,
\Big(\!\stackrel1{U}\!(x,y;v)\!
\stackrel2{U}\!(x'\!,y;w)\;\O_\gk
\stackrel2{U}\!(y,y';w)\Big)\;.\nn
\ea
with $\th$ from \Ref{theta} and
\ben
f(x;v,w)\equiv 
\frac{1\!-\!2\sr(x;w)\sr(x;v)\!+\!\sr^2(x;w)}{1\!-\!\sr^2(x;w)}\;.
\een

Let us recall that the limits of these expressions for the
corresponding flat space models do not exist as the result depends on
the order in which the limits are taken due to the different
coefficients of the $\th$ functions \ci{VeEiMa84}. For the
gravitationally coupled models, however, these ambiguities disappear
altogether by virtue of the coordinate dependence of the spectral
parameter if we have $\p_w \sr(\r,\rt;w)=0$ at the limit points
\ci{KorSam98}!  This is indeed the case with appropriate boundary
conditions on the dilaton and its axionic partner as we have already
seen in \Ref{limits}.

Thus, \Ref{PBU} yields a well-defined algebra for the modified
transition matrices $\Ut$ connecting two of these critical
points $x_0$, $y_0$. With 
\be\la{Uc}
U(w)\equiv \Ut(x_0,y_0;w)\;,
\ee
the final Poisson algebra takes the form
\be\la{UU}
\left\{\stackrel1{U}\!(v)\;,\,\stackrel2{U}\!(w)\right\}~=~ 
\left[\frac{2\O_\eee}{v-w}\,,\, 
\stackrel1{U}\!(v)\stackrel2{U}\!(w)
\right]\;,
\ee
for coinciding signs of the imaginary parts of $v$ and $w$. The
Poisson brackets between transition matrices for $v$ and $w$ from
different halves of the complex planes depend on the concrete behavior
of $\r$ and $\rt$ in the end-points. They may e.g.\ coincide with
\Ref{UU} or with \Ref{U2} below.

In contrast, the Poisson structure of the monodromy matrix $\cM$ is
universal and may be computed from \Ref{PBU} and the definition
\Ref{M} to 
\ba\la{PBM}
\left\{\stackrel1{\cM}\!(v)\;,\,\stackrel2{\cM}\!(w)\right\}
&=&  \frac{2\O_\eee}{v-w} \stackrel1{\cM}\!(v)\stackrel2{\cM}\!(w)\,
+  \stackrel1{\cM}\!(v)\stackrel2{\cM}\!(w) \,\frac{2\O_\eee}{v-w} \\
&& {} -
\stackrel1{\cM}\!(v)\,\frac{2\O_\eee^\t}{v-w}\stackrel2{\cM}\!(w)
\,- \stackrel2{\cM}\!(w)\,\frac{2\O_\eee^\t}{v-w}\stackrel1{\cM}\!(v) 
\;,\nn
\ea
with $\O_\eee$ and $\O_\eee^\t$ from \Ref{O1} and \Ref{O2},
respectively. One may check that indeed these brackets are compatible
with the symmetry \Ref{Msym}, as required for consistency.  For the
purpose of quantization and representation of \Ref{PBM} it is further
convenient to decompose this structure according to \Ref{Upm} into the
following brackets
\ba
\left\{\stackrel1{U_\pm}\!(v)\;,\,\stackrel2{U_\pm}\!(w)\right\}
&=& \left[\frac{2\O_\eee}{v-w}\,,\, 
\stackrel1{U_\pm}\!(v)\stackrel2{U_\pm}\!(w)
\right]\;,\la{U1}\\
\left\{\stackrel1{U_\pm}\!(v)\;,\,\stackrel2{U_\mp}\!(w)\right\}
&=& \frac{2\O_\eee}{v-w} \stackrel1{U_\pm}\!(v)\stackrel2{U_\mp}\!(w)
-\stackrel1{U_\pm}\!(v)
\stackrel2{U_\mp}\!(w)\frac{2\O_\eee^\t}{v-w}\;,
\la{U2}\ea
which may be easier to handle due to the similarity of \Ref{U1} with
the well-known Yangian algebra $Y\!(\eee)$ \ci{FaSkTa79,Drin85,Bern91}.    

\mathversion{bold}
\subsection{Hidden symmetries: The Lie-Poisson action of $\9$}
\mathversion{normal}
\la{LP}

Previous studies of the Geroch group and its generalizations have been
mostly concerned with the non-linear and non-local realization of
these groups on the physical fields and their dual potentials at the
level of the equations of motion (recall that duality symmetries in
even dimensions are always on-shell).  Now, for the flat space
$\s$-models it has been known for a long time that this action is not
symplectic \ci{LuePoh78,DHLM82}. Rather, it represents a Lie-Poisson
action of the associated symmetry group \ci{Seme85,BabBer92}. Here, a
similar picture emerges, since the hidden symmetries of dimensionally
reduced supergravity can be recovered via a Lie-Poisson action of
$\mathfrak{e}_9$, which is canonically generated by the integrals of
motion $U(w)$ from \Ref{Uc} \ci{KorSam97a,KorSam98}.\footnote{See e.g.
  \ci{BelZak78,BreMai87,BerJul97} for discussions of the Geroch 
  group within the general framework of dressing transformations.}

In this section, we show how this result fits into the canonical
framework established in the foregoing sections.  In
particular, we give the action of $\eeee$ on all the fermionic fields
involved in the model.  Our basic objects are the transition matrices
$U(w)$ from \Ref{Uc}, where we assume that $x_0$ and $y_0$ are the
spatial boundaries and critical in the sense discussed above, namely
that $\Lt_0$ vanishes at these points. In particular, the transition
matrix $U(t,x_0,x;w)$ then provides a solution of the linear system
$\Ref{ls}$ which we denote by $\cVh_0$.

We define the following matrix-valued symmetry generator
\be\la{symmetry0}
G(v)~\equiv~
\ad_{\,U(v)}U^{-1}(v)\;,
\ee
where ``$\ad$'' denotes the adjoint action via the canonical Poisson
structure. In matrix components the action of \Ref{symmetry0} on an
arbitrary phase space function $f$ reads
\ben
G^{ab}(v)\:f ~\equiv~ \left\{U^{ac}(v)\,,f\right\}
\left(U^{-1}(v)\right)^{cb}\;,
\een
with $\E$ indices $a, b, \dots$ (cf.\ Appendix
\ref{Aten}). Making use of \Ref{PBU} we can determine the action of
\Ref{symmetry0} on the monodromy matrix $\cM$:
\be
\stackrel1{G}\!\!(v)\,\stackrel2{\cM}\!(w)~=~ 
\frac{1}{v\!-\!w}\;\left(\O_\eee\stackrel2{\cM}\!(w)\,
- \stackrel2{\cM}\!(w)\,\O_\eee^\t\right)\;.
\ee
This motivates the definition of the following symmetry operator
\be\la{symmetry}
G[\L] ~\equiv~ 
\oint_{\ell}\,\frac{dw}{2\pi\i}\;\tr\!\left(\L(w)\,G(w)\right)\;,
\ee
parametrized by an algebra-valued function $\L(w)\!\in\!\eee$, regular
along the real $w$-axis and vanishing at $w\!\ra\!\8$. The path
$\ell$ is chosen to encircle the real $w$-axis, such that $\L(w)$ is
holomorphic inside the enclosed area. The monodromy matrix transforms
under the action of $G[\L]$ as
\be\la{actionM}
G[\L]\,\cM(w) ~=~ \L(w)\cM(w) - \cM(w)\,\t(\L(w))\;.
\ee
The orbit of the monodromy matrix under the symmetry group fills the
complete set of $E_8$-valued functions with symmetry \Ref{Msym} and
the assumed analyticity properties on the real axis. The symmetry
group thus acts transitively if the monodromy matrix parametrizes the
full phase space (which e.g.\ is the case for cylindrical
gravitational waves).

With the general formula
\be\la{TX}
\left\{U(x,y;v)\,,f\,\right\}
=\int_x^y dx'U(x,x';v)\left\{
L_1(x'\!,\sr(x'\!,v))\,,f\,\right\}
U(x'\!,y;v) \;,
\ee
and the fundamental Poisson brackets \Ref{PBb1}--\Ref{PBbf2} we can
directly compute the symmetry action on the physical fields. It turns
out that the relevant parameter which describes this action is the
combination 
\be\la{lhat}
\cVh_0^{-1}\L \cVh_0 ~\equiv~
\ft12 \Lh^{IJ} X^{IJ} + \Lh^A Y^A\;,
\ee
(note that $\Lh$ depends on both $w$ and the fields, whereas
$\L(w)$ is coordinate independent). The symmetry action on the bosonic
matrix $\cV(x)$ becomes  
\ba
G[\L]\, \cV(x) &=& \oint_\ell \frac{dw}{2\pi\i}\; \left(
\frac{2\sr}{\r(1\!-\!\sr^2)}\,\cV(x)
\,\Lh^A(\sr(w)) Y^A\right)
\la{actionV}\\
&=& - \oint_{\sr(\ell)} \frac{d\sr}{2\pi\i\sr}\;\left( \cV(x)
\,\Lh^A(\sr) Y^A\right)
\;.\nn
\ea

The action on the fermionic fields of the model is given by
\ba \la{E9fermions}
G[\L]\, \psi_{2\,\pm}^{I} &=& 
- \oint_\ell\frac{dv}{2\pi\i}\,\left(\frac{4\sr}{\r(1\!\pm\!\sr)^2}\,
\Lh^{IJ}(\sr)\psi_{2\,\pm}^J \right)\;,\la{actf}\\
G[\L]\, \chi_\pm^{\dot{A}} &=& 
-\oint_\ell\frac{dv}{2\pi\i}\,\left(\frac{\sr}{\r(1\!\pm\!\sr)^2}\,
\G^{IJ}_{\dot{A}\dot{B}}\,\Lh^{IJ}(\sr)\chi^{\dot{B}}_\pm +
\frac{8\sr^2}{\r(1\!\pm\!\sr)^2(1\!-\!\sr^2)}\,
\G^I_{A\dot{A}}\,\Lh^{A}(\sr)\psi_{2\,\pm}^I
\right),\non
G[\L]\, \psi_{\pm}^{I} &=&
-\oint_\ell\frac{dv}{2\pi\i}\,\left(\frac{4\sr}{\r(1\!\pm\!\sr)^2}\,
\Lh^{IJ}(\sr)\psi_{\pm}^J 
+\frac{16\sr^2}{\r(1\!\pm\!\sr)^4}\,
\Lh^{IJ}(\sr)\psi_{2\,\pm}^J \right)\non
&& \pm  \oint_\ell\frac{dv}{2\pi\i}\,\left(
\frac{8\sr^2}{\r(1\!\pm\!\sr)^2(1\!-\!\sr^2)}\,
\G^I_{A\dot{B}}\,\Lh^{A}(\sr) \chi^{\dot{B}}_\pm \right)\;.\nn
\ea
These transformations preserve the chirality of the fermionic
fields in the following sense. The equations of motion \Ref{feqm}
admit chiral solutions, i.e. solutions with either all $+$ or all $-$
fermionic components switched off. The action \Ref{actf} then takes
place within these sectors. The respective first terms in \Ref{actf}
are pure $\SO$ gauge transformations \Ref{so16} when acting on a
chiral solution. In general, they can not be absorbed by this gauge
freedom owing to the different coefficients for the chiral halves.
Observe also that the transformations preserve the super-Weyl gauge,
where $\psi^I_2=0\,$.\footnote{In this gauge, the formula for the
  variation of $\chi^{\dot A}_\pm$ coincides with (4.2.29) of
  \ci{Nico91}, which is obtained from the particular values
  $\Upsilon_{\!\L}(x,\sr\!=\!\mp1)$ of the compensating
  $\gh^\8$-rotation from \Ref{actionVh} below.}

The homogeneous action of the symmetry on the fermionic
fields has the important consequence that it cannot act transitively
on the space of all solutions. It is not possible to generate
fermionic from purely bosonic solutions of the classical equations of
motion in this way.

The action on the conformal factor $\s$ is given by
\be\la{actions}
G[\L]\,\s ~=~ \oint_\ell\, \frac{dw}{2\pi\i}\;
\tr\left( \L\p_w\cVh_0\cVh_0^{-1}\right)\;,
\ee
in agreement with the result derived in \ci{Nico91}. Formula
\Ref{actions} is easily obtained from 
\ben
G(w)\:\p_1\s(x) ~=~ 
-\cVh_0(x,\sr(w))\,\p_w L_1(x,\sr(w))\, \cVh_0^{-1}(x,\sr(w))\;,
\een
which in turn follows from \Ref{PBb1a}, \Ref{TX} and the fact that
$\p_w L_1=\p_{\rt} L_1$ (cf.\ \Ref{sr}). 

Finally, we can also give the transformation behavior of the solution
$\cVh_0$ of the linear system. Evaluating \Ref{PBU} we obtain
\be\la{actionVh}
G[\L]\,\cVh_0(x,\sr(w)) ~=~
\L(w)\cVh_0(x,\sr(w))-
\cVh_0(x,\sr(w))\,\Upsilon_{\!\L}(x,\sr(w))\;, 
\ee
where
\ba
\Upsilon_{\!\L}(x,\sr(w))&=&
\oint_{\ell}\:\frac{dv}{2\pi\i(v\!-\!w)}\;
\ft12\Lh^{IJ}X^{IJ} \non
&&{}+~\frac{1\!-\!\sr^2(w)}{\sr(w)}\;
\oint_{\ell}\:\frac{dv}{2\pi\i(v\!-\!w)}\;
\frac{\sr(v)}{1\!-\!\sr^2(v)}\;
\Lh^AY^A \;,\nn
\ea

The matrix $\L(w)$ depends on the constant spectral parameter
$w$; whereas $\Upsilon_{\!\L}(x,t,\sr(w))$ depends on the
variable spectral parameter $\sr$ and satisfies
\be\la{cond}
\Upsilon_{\!\L}(x,\sr(w)) ~=~ 
\t^\infty\!\left(\Upsilon_{\!\L}(x,\sr(w))\right)~=~
\t\!\left(\Upsilon_{\!\L}(x,\sr^{-1}(w))\right)\;.
\ee

This result provides the link to the traditional realization of the
Geroch group via the linear system. There, the space of classical 
solutions is formally identified with the infinite dimensional 
coset space $\GG^\8/\,\GH^\8$, where the underlying algebra $\gg^\8$ 
is generated by $\eee$ valued functions in the $w$-plane while 
$\gh^\8$ is defined as the set of $\eee$ valued functions in the 
$\sr$-plane satisfying \Ref{cond}. Like in \Ref{symmetry} the 
symmetry algebra is parametrized by a function $\L(w)\!\in\!\gg^\8$. 
Starting from a solution $\cVh_0(\sr)$ which is holomorphic in the
unit disc $|\sr|\!\leq\!1$, the function
$\Upsilon_{\!\L}\!\in\!\gh^\8$ in \Ref{actionVh} is then uniquely
defined so as to restore this holomorphy which is violated by the
pure action of $\L(w)$. Indeed, it follows from the form of
$\Upsilon_{\!\L}$ and \Ref{lhat}, that in \Ref{actionVh} the r.h.s.\ 
multiplication of $\cVh_0$ with $\Upsilon_{\!\L}$ removes all
singularities caused by the l.h.s.\ multiplication with $\L(w)$ from
the unit disc (note that the path $\ell$ surrounds the unit disc in
the $\sr$-plane).  

We close this section with some remarks on the algebraic structure of
the symmetry. The canonical realization given in \Ref{symmetry} is
parametrized by meromorphic $\eee$-valued functions and yields closed
expressions for the action on the physical fields.  The algebra of
these operators is most conveniently obtained from
\Ref{actionM}, which immediately shows
\begin{equation}\la{affine}
\Big[G[\L_1],G[\L_2]\Big] ~=~ G\Big[[\L_1,\L_2]\Big]\;.
\end{equation}

Half of the affine algebra $\eeee$ may be recovered by formal Laurent
expansion around $w\!=\!\8$:~\footnote{There is a slight subtlety here,
  since strictly speaking the functions $\L(w)\!=\!\L_n w^n$ do not
  belong to the class of functions for which we have defined
  \Ref{symmetry}. Since the integrand is singular at infinity,
  definition \Ref{symmetry} depends on the precise choice of the
  contour in this region, which has not been specified above.}
\be\la{exL}
\L(w)~=~I+\L_1w+\L_2w^2+\dots\;,
\ee
The action of these modes on the physical fields follows from
expansion of the respective closed formulas \Ref{actionV}--\Ref{actf}.
In a rather formal sense \Ref{exL} may be related to the expansion 
\be\la{exG}
G(w)~=~I+\frac1{w}\;G_1+\frac1{w^2}\;G_2+\dots\;,
\ee
obtained from \Ref{symmetry0} by handling the linear system \Ref{ls}
as a formal power series in $w^{-1}$. Observe, however, that
$G(w)$ does not allow a Laurent expansion around $w=\8$
when the branch point moves to infinity.

The conventional realization of $\eeee$ also includes a representation
of the other half of the affine algebra; this is achieved by the
introduction of infinitely many additional gauge degrees of freedom
associated with the ``maximal compact subgroup'' $\GH^\8$ of $\9$
\ci{BreMai87,JulNic96}. Furthermore, this group possesses a central
extension which acts trivially on the physical fields but shifts the
conformal factor $\s$ by a constant \ci{Juli83}.  Although so far
there does not exist a canonical formulation with an enlarged phase
space capturing these extra symmetries, the proper action of the
central term is implied by formula \Ref{actions}, which itself is
canonically generated.

The form of the symmetry operators \Ref{symmetry0} shows that this
action is not symplectic but satisfies
\be\la{LieP}
G(w)\left\{f_1,f_2\right\} =
\left\{G(w)f_1,f_2\right\} + \left\{f_1,G(w)f_2\right\} +
\left[G(w)f_1\,,G(w)f_2\right]\;,
\ee
on any two phase space functions $f_1$, $f_2$, where the commutator on
the r.h.s.\ is understood for the matrix-valued action of $G(w)$.
This is an example of a Lie-Poisson action, i.e.\ it does not preserve
the Poisson structure on the phase space but on the direct product of
the phase space with the symmetry group \ci{Seme85,BabBer92}. This fact
becomes crucial upon quantizing the structure, since it is not $\eeee$
but its quadratic deformation underlying \Ref{UU} according to the
representations of which the spectrum of physical states will have to
be classified.

\section{Outlook}
Our ultimate interest in developing the canonical framework is in
quantizing $\216$ supergravity, and this not only in view of 
constructing exactly solvable and ``sufficiently complicated'' models
of matter coupled quantum gravity in two dimensions. We are equally
motivated by recent developments in non-perturbative string theories,
where (finite dimensional and discrete) duality symmetries play a
central role.

In standard canonical quantization one converts the Poisson (or Dirac)
brackets of the basic fields into (anti)commutators and the constraints
into operators acting in a suitable Hilbert space. Irrespective of
the possible ambiguities in this procedure, one must
ensure that these operator constraints are indeed well defined 
in the sense that their matrix elements exist between any two 
states; this may require some ``renormalization'', such as normal
ordering in string theory. The next step would be to search for
physical states which by definition are annihilated by the quantum
supersymmetry generators. This is a difficult task because $\16$
supergravity is a fully interacting theory on the worldsheet, unlike
the conformal supergravities giving rise to superstring theories.
If the classical constraint algebra can be transferred to the quantum 
theory (possibly modulo certain anomalies), the supersymmetry algebra
would ensure that the physical states are also annihilated by the
Hamiltonian (Wheeler-DeWitt operator), as well as the diffeomorphism
and $\SO$ constraints.

Now, it would seem rather foolhardy to hope to be able to carry through 
such an ambitious program for the Lagrangian \Ref{L2} directly.  However,
the integrability of the model, which is encapsulated in the existence
of infinitely many conserved non-local charges, may open an
alternative representation theoretic route to its quantization.
Clearly, the physical states should transform as a representation of
the (quantum) algebra generated by the conserved charges. In this way,
these charges would give rise to a kind of spectrum generating algebra
akin to the one of string theory, which is generated by the DDF
operators. To properly work out these ideas may involve an extension
of the techniques that have been successfully applied to the
computation of form factors of the flat space non-linear ${SO}(3)$
$\s$-model \ci{Smir92}.\footnote{Another object of interest in flat
  space integrable quantum field theories is the exact $S$-matrix.
  However, in a theory of quantum gravity, the very notion of an $S$-matrix 
  is a priori meaningless, except in those special circumstances 
  corresponding to asymptotically flat spacetimes which allow for asymptotic 
  states to exist.} The crucial difference with the flat space theories 
is the requirement of consistency with local supersymmetry.

According to the discussion in section \ref{LP}, and contrary to one's
naive expectations the physical states cannot be expected to fall into
standard $\eeee$-multiplets because the action of $\eeee$ is not
symplectic. Even if this were the case, we note that almost nothing is
known about the unitary irreducible representations of $\9$; even for
its finite dimensional subgroup $\EE$, the representation theory is
still only rudimentary.  At any rate, a minimum requirement for
analyzing the physical spectrum will thus be to find the quantum
algebra underlying the classical Poisson algebras \Ref{PBM} and
\Ref{U1}, \Ref{U2}.  These structures are closely related to the
Yangian algebra $Y\!(\eee)$ \ci{FaSkTa79,Drin85,Bern91}, however with
different analyticity properties, the novel ``twist'' structure of
\Ref{U2} and the additional symmetry \Ref{Msym}. Quantization must
respect these properties.  For the models with coset space
$SL(N,\R)/SO(N)$ the problem of directly quantizing \Ref{PBM} has been
solved in \ci{KorSam98} by use of several well-known results on the
corresponding Yangians. Although $\E$ is a much more complicated
group, it turns out that the corresponding $R$-matrix has already been
worked out in the mathematical literature \ci{ChaPre91}.  Curiously,
this $R$-matrix exists only if an extra singlet is added to the $\bf
248$ of $\E$; in other words, the quantum $U$-matrix must be extended
by one row and one column to a 249 by 249 matrix if it is to obey all
consistency conditions. This seems to indicate that the full quantum
symmetry will involve additional degrees of freedom from the
gravitational sector.

The quantization of $\16$ supergravity may resolve various puzzles
related to the presence of fermions. While the Geroch group and its
analogs are known to act transitively on the space of classical
bosonic solutions (modulo some technical assumptions), we do not know
how to generate fermionic solutions from purely bosonic ones (the
vacuum, in particular), see the remarks accompanying \Ref{E9fermions}.
This would require a superextension of $\9$.  In principle, local
supersymmetry can give rise to global supercharges, but only in very
special backgrounds admitting Killing spinors.  Performing a Geroch
transformation on such a background will destroy this property in
general. This makes the existence of a superalgebra containing $\eeee$
as its maximal bosonic subalgebra somewhat unlikely.  A further
indication that something is amiss here is that a supersymmetric
generalization of the Breitenlohner-Maison cocycle formula
\ci{BreMai87} for the conformal factor has so far not been found.
Lastly, it is not clear what physical significance should be attached
to classical solutions depending on anticommuting $c$-numbers. In the
quantum theory, all these problems may dissolve by themselves. The
first because the quantum theory may not admit purely bosonic physical
states, like simple exactly solvable models of quantum supergravity in
three dimensions \ci{WiMaNi93} or perturbatively treated canonical
$N\!=\!1$ quantum supergravity in four dimensions \ci{CFOP94}. The
others because the fermions become operators, and only expectation
values of observables (such as the conserved charges) are physically
meaningful quantities in quantum gravity.

Finally, the Geroch group and its generalizations represent infinite 
dimensional extensions of the duality symmetries that have played such 
an important role in recent developments of string theory. We have not
yet studied the case of bounded ``open'' worldsheets and the effect of
these symmetries on the boundary conditions (which in the case of open
strings have lead to the discovery of D branes \ci{Polc96}). 
A first step in this direction would be to find out whether
the Geroch group can be implemented with periodic boundary
conditions on the space coordinates. Unlike $T$-duality, which
simply involves an interchange of $x^0$ and $x^1$, and thus
of Neumann and Dirichlet boundary conditions, we here face 
the challenge of making an infinite number of duality
transformations compatible with the boundary conditions.
{}\\
{\bf Acknowledgments:} H.\,S. is grateful to Studienstiftung
des deutschen Volkes for financial support. H.\,N. has benefitted
from discussions with B.~Julia on Lie-Poisson structures, and would 
like to thank B.~Zumino for pointing out the possible relevance of 
Yangians for dimensionally reduced gravity already some time ago.
We wish to thank K.~Koepsell for discussions on $Y\!(\eee)$.

\bigskip
\begin{appendix}
\section{Conventions}
\subsection{Metric and spinor conventions}\la{Aspin}
Throughout this paper we work with the flat metric 
$\eta_{\a\b} = \rm{diag}(+-)$. For any vector $V_\a$ and covector
$V^\a$, we define the light-cone components  
\be
V_\pm := \ft12(V_0\pm V_1) \;,
\qquad V^\pm: = V^0\pm V^1 = V_0 \mp V_1\;,
\ee
respectively, so that $V^\a W_\a = V_+ W_- + V_-W_+ $. 

The $\g$-matrices obey
\be
\g_\a\g_\b = \eta_{\a\b} + \e_{\a\b}\g^3 \;, \quad
\g^3\g_\a = \e_{\a\b}\g^\b\;,\quad\g^3\g_\pm = \mp \g_\pm
\ee
with $\e_{01}=-\e^{01} = 1$, and are explicitly given by
\be
\g_0=\left({0\atop\i}\;\,{-\i\atop0}\right)\;,\quad
\g_1=\left({\!\!-\i\atop0}\;\,{\;0\atop\i}\right)\;,\quad
\g^3=\left({0\atop1}\;\,{1\atop0}\right)\;.
\ee
We thus make use of the Majorana  representation where the
charge conjugation matrix is $\cC=\g_0$, such that a 
Majorana spinor obeying $\overline{\psi}=\psi^T\cC$ has
two real components. Any Majorana Weyl spinor decomposes as 
\be
\ft12(1\pm\g^3)\psi~\equiv~
\left(\psi_\pm\atop{\pm\psi_\pm}\right)
\quad\Longrightarrow\quad
\g_\pm (1\mp \g^3) \psi = 0 \;.
\ee
The one component spinors $\psi_\pm$, etc. are thus to be
treated as real anticommuting variables.
Let us also give some useful rules for the transcription between 
two component and one component notation:
\ba
\overline{\psi}\chi &=& 2\i(\psi_+\chi_- - \psi_-\chi_+)\;,  \qquad
\overline{\psi}\g^3\chi = -2\i(\psi_+\chi_- +\psi_-\chi_+)\;, \non
\overline{\psi}\g_+\chi &=& 2\psi_+\chi_+ \;,  \qquad
\overline{\psi}\g_-\chi = 2\psi_-\chi_-\;. \nn
\ea

\subsection{$E_8$ Conventions}\la{Ae8}

Under its $\SO$ subgroup, the fundamental (=adjoint) representation of
$\EE$ decomposes as ${\bf 248}\ra {\bf 120}\oplus{\bf 128}$.  We
denote the 120 generators of the maximal compact subgroup $\SO$ by
$X^{IJ}=-X^{JI}$ with $\SO$ vector indices $I, J, \dots=1,\dots,16$,
and the non-compact generators by $Y^A$, where $A, B, \dots$ (and
$\dot A, \dot B, \dots$) $=1, \dots, 128$ label the left (right)
handed spinor representation of $\SO$.  The defining relations for the
Lie algebra $\eee$ are
\ba
{}\left[X^{IJ},X^{KL}\right] &=& 
\d^{JK}X^{IL} - \d^{IK}X^{JL} + \d^{IL}X^{JK} - \d^{JL}X^{IK}
\;, \la{e8}\\
{}\left[X^{IJ},Y^A\right] &=& -\ft12\G^{IJ}_{AB}\,Y^B \qquad
{}\left[Y^A,Y^B\right] = \ft14\G^{IJ}_{AB}\,X^{IJ} \;,\nn
\ea
where the $\G^{IJ}_{AB}$ denote the $SO(16)$-$\G$-matrices which fulfill 
\be\la{Gamma}
\G^I_{A\dot{A}}\G^J_{\dot{A} B} = \d^{IJ}_{AB} + \G^{IJ}_{AB}\;.
\ee

In the adjoint representation the generators are normalized such that
\ben
\tr\left(X^{IJ}X^{KL}\right) = -120 \,\d^{IJ}_{KL}\;,\qquad
\tr\left(Y^AY^B\right) = 60 \,\d^{AB} \;.
\een
For the current 
\ben
\cV^{-1}\p_\m\cV = \ft12 Q^{IJ}_\m X^{IJ} + P^A_\m Y^A\;,
\een
this yields
\ba
Q^{IJ}_\m = -\ft1{60} \tr \left(X^{IJ}\cV^{-1}\p_\m\cV\right)\;,\qquad
P^A_\m = \ft1{60} \tr \left(Y^A\cV^{-1}\p_\m\cV\right) \;.\nn
\ea

We denote the splitting of $\eee$ into its compact and non-compact
part as 
\be\la{hk}
\eee ~=~ \so \oplus \gk\;,
\ee
where $\gk$ is (as a vector space) generated by the $Y^A$. This
splitting defines the involutive algebra-automorphism $\t$
\be\la{tau}
\t\left(X^{IJ}\right)=X^{IJ}\;,\quad \t\left(Y^{A}\right)=-Y^{A}\;.
\ee
By exponentiation this involution is lifted to the group $E_8$.

\subsection{Tensor conventions}\la{Aten}
Finally, we explain the index-free tensor notation for $\E$, in terms
of which some formulas in the main text can be cast into a more
compact form. We consider the $248$-dimensional adjoint matrix
representation of $\eee$, on which by integration also $\E$ is
represented. We label the matrix indices by $a, b, c, \dots=1, \dots,
248$.

For any matrix $A^{ab}$ we define the corresponding matrices
acting in the tensor product of two representation spaces:
\ben
\stackrel{1}{A}\;\equiv A\otimes I\qquad\mbox{and}\qquad
\stackrel{2}{A}\;\equiv I\otimes A\;.
\een
In components this takes the form
  $(A\otimes I)^{ab,cd} \equiv A^{ab}\d^{cd}$
  and $(I\otimes A)^{ab,cd} \equiv
  \d^{ab}A^{cd}\;$. 
Following \ci{FadTak87} we then introduce the matrix notation
for Poisson brackets:  
\be\la{free}
\Big\{\stackrel1{A}\;,\,\stackrel2{B}\Big\}^{ab,cd} ~\equiv~
\{A^{ab},B^{cd}\}\;,
\ee
for matrices $A^{ab}$, $B^{cd}$.
In this notation the canonical brackets \Ref{PBc} become
\ben
\Big\{\,\stackrel1{\cV^{-1}\p_1\cV}\!(x)\;,\,
\stackrel2{\Pi}\!(y)\Big\} ~=~ \O_\eee\, \d(x\!-y)\;,
\een
where the Casimir element $\O_\eee$ of $\eee$ is defined as
\be\la{O1}
\O_\eee~\equiv~\O_\so + \O_\gk ~\equiv~ 
     \ft1{2}\, X^{IJ}\!\otimes\!X^{IJ} 
     - Y^A\!\otimes\!Y^A ~\in~ \eee\otimes\eee\;.
\ee
In addition, we need the following ``twisted'' Casimir element which
appears in the Poisson brackets \Ref{PBM}, \Ref{U2}:
\be\la{O2}
\O^\t_\eee~\equiv~\O_\so - \O_\gk ~\equiv~ 
  \ft1{2}\, X^{IJ}\!\otimes\!X^{IJ} 
   +Y^A\!\otimes\!Y^A ~\in~ \eee\otimes\eee\;.
\ee

\end{appendix}


\end{document}